\documentclass[12pt]{JHEP3}
\usepackage{amssymb}
\usepackage{amsmath}
\usepackage{amsfonts}
\usepackage{graphicx}
\newcommand\be{\begin{eqnarray}}
\newcommand\ee{\end{eqnarray}}

\newcommand\e{\epsilon}

\newcommand\half{\frac{1}{2}}
\renewcommand\d{\partial}

\newcommand{\x}[1]{\text{#1}}

\newcommand{\gSU}[1]{\mathop{\rm SU}(#1)}
\newcommand{\gSO}[1]{\mathop{\rm SO}(#1)}

\newcommand{\gU}[1]{\mathop{\rm {}U}(#1)}
\newcommand{\grp}[2]{\mathop{\rm #1}(#2)}

\newcommand{\wt}{\widetilde}

\newcommand{\idn}{{1\relax{\kern-.35em}1}}

\newcommand{\Rf}{\mathbb{R}}
\newcommand{\Zf}{\mathbb{Z}}

\newcommand{\Nc}{\mathcal{N}}

\DeclareMathOperator{\tr}{Tr}

\newcommand{\gym}{g_{\x{YM}}}

\author{Ofer Aharony$^{a,b}$, Zohar Komargodski$^a$, Assaf Patir$^a$\\
$^a$Department of Particle Physics, Weizmann Institute of Science,
Rehovot 76100, Israel\\
$^b$SITP, Department of Physics and SLAC, Stanford University,
Stanford, CA 94305, USA\\ E-mail:
\email{ofer.aharony@weizmann.ac.il},
\email{zohar.komargodsky@weizmann.ac.il},
\email{assaf.patir@weizmann.ac.il}}

\abstract{We discuss the moduli space of nine dimensional $\Nc=1$
supersymmetric compactifications of M theory / string theory with
reduced rank (rank $10$ or rank $2$), exhibiting how all the
different theories (including M theory compactified on a Klein
bottle and on a M\"obius strip, the Dabholkar-Park background, CHL
strings and asymmetric orbifolds of type II strings on a circle) fit
together, and what are the weakly coupled descriptions in different
regions of the moduli space. We argue that there are two
disconnected components in the moduli space of theories with rank
$2$. We analyze in detail the limits of the M theory
compactifications on a Klein bottle and on a M\"obius strip which
naively give type IIA string theory with an uncharged orientifold
8-plane carrying discrete RR flux. In order to consistently describe
these limits we conjecture that this orientifold non-perturbatively
splits into a D8-brane and an orientifold plane of charge $(-1)$
which sits at infinite coupling. We construct the M(atrix) theory
for M theory on a Klein bottle (and the theories related to it),
which is given by a $2+1$ dimensional gauge theory with a varying
gauge coupling compactified on a cylinder with specific boundary
conditions. We also clarify the construction of the M(atrix) theory
for backgrounds of rank $18$, including the heterotic string on a
circle.}

\preprint{WIS/02/07-FEB-DPP}


\title{The Moduli Space and M(atrix) Theory of 9d $\Nc=1$
Backgrounds of M/String Theory}
\begin{document}
%%%%%%%%%%%%%%%%%%%%%%%%%%%%%%%%%%%%%%%%%%%%%%%%%%%%%%%%%%%%%%%%%%%%%%%%%%%%%
\section{Introduction}

In this paper we discuss in detail the structure of the moduli
space of nine dimensional $\Nc=1$ supersymmetric backgrounds of M
theory and string theory, and their M(atrix) theory construction.
There are two main motivations for this study :
\begin{itemize}
    \item The global structure of the moduli space of (maximally
supersymmetric) toroidal compactifications of M/string theory has
been studied extensively, and all of its corners have been mapped
out. Less is known about compactifications which preserve only half
of the supersymmetry ($16$ supercharges).\footnote{Some of these
backgrounds were studied in \cite{deBoer:2001px}.} The moduli space
of such nine dimensional backgrounds with rank $18$ (including
heterotic strings on a circle) has been extensively studied, and all
its corners are understood; however it seems that no similar study
has been done for backgrounds with lower rank (rank $10$ or rank
$2$). These backgrounds include the compactification of M theory on
a Klein bottle and on a M\"obius strip. In this paper we study the
moduli space of these backgrounds in detail. We will encounter two
surprises : we will see that the moduli space of backgrounds of rank
$2$ has two disconnected components,\footnote{This was independently
discovered by A. Keurentjes \cite{Keur-private}.} and we will see
that both for rank $2$ and for rank $10$ there is a region of the
moduli space which has not previously been explored, and whose
description requires interesting non-perturbative corrections to
some orientifold planes in type IIA string theory.
    \item The theories we discuss (with rank $n$) have non-trivial
duality groups of the form $\grp{SO}{n-1,1,\Zf}$. In their
heterotic descriptions these are simply the T-duality groups.
However, from the M theory point of view these dualities are quite
non-trivial. In particular, it is interesting to study the
manifestation of these duality groups in the M(atrix) theory
description of these backgrounds; T-duality groups in space-time
often map to interesting S-duality groups in the M(atrix) theory
gauge theories. In this paper we will derive the M(atrix) theory
for some of these backgrounds; the detailed discussion of the
realization of the duality in M(atrix) theory is postponed to
future work.
\end{itemize}

We begin in section \ref{backgrounds} with a detailed discussion
of the structure of the moduli space of 9d $\Nc=1$ backgrounds. We
review the structure of the moduli space of rank $18$ backgrounds,
since this will have many similarities to the moduli spaces of
reduced rank, and we then discuss in detail all the corners of the
moduli spaces of reduced rank. In section \ref{Ms} we construct
the M(atrix) theory for the backgrounds corresponding to M theory
on a cylinder (with a specific light-like Wilson line) and on a
Klein bottle. The case of the cylinder has been constructed before
\cite{Kabat:1997za}, but we clarify its derivation and the mapping
of parameters from space-time to the M(atrix) theory. Our
construction for the Klein bottle is new. We end in section
\ref{Conclusions} with our conclusions and some open questions.
Four appendices contain various technical details.

\section{The moduli space of nine dimensional $\Nc=1$ backgrounds}
\label{backgrounds}

In this section we review the moduli space of compactifications of
string/M theory to nine dimensions which preserve $\Nc=1$
supersymmetry. On general grounds, if such a compactification has
a rank $n$ gauge group in its nine dimensional low-energy
effective action, its moduli space takes the form
$\gSO{n-1,1,\Zf}\backslash \gSO{n-1,1}/\gSO{n-1}\times \Rf$, with
the first component involving $n-1$ real scalars sitting in $n-1$
vector multiplets and the second component involving the scalar
sitting (together with an additional $\gU1$ vector field) in the
graviton multiplet. In different regions of this moduli space
there are different weakly coupled descriptions of the physics. We
begin by reviewing the $n=18$ case which is well-known. We then
discuss the cases of $n=2$ and $n=10$, where we will encounter
some surprises and some regions which have not previously been
analyzed.

\subsection{A review of nine dimensional $\Nc=1$ theories with rank $18$}\label{rank16review}

In this subsection we review the different regions of the moduli
space of nine dimensional compactifications of M theory and string
theory with 16 supercharges and a rank 18 group, including M theory
on $\Rf^9\times S^1\times(S^1/\Zf_2)$, the two heterotic strings
compactified on a circle, the type I string on a circle and the type
I' string. The results are mainly from \cite{Horava:1995qa} and
\cite{Polchinski:1995df}. The full moduli space for these theories
is $\gSO{17,1,\Zf}\backslash\gSO{17,1}/\gSO{17}\times\Rf$.
Naturally, different descriptions are valid in different regions of
this space. For simplicity, we shall restrict our discussion to the
subspaces of moduli space that have enhanced $E_8\times E_8$ and
enhanced $\gSO{32}$ gauge symmetries at low energies. These
subspaces take the form $\gSO{1,1,\Zf}\backslash \gSO{1,1}\times
\Rf$, so they are analogous to the rank 2 case, which is the main
subject of this paper, and we will see many similarities between
these two cases. However, a reader that is familiar with rank 18
compactifications is welcome to skip to the next subsection.

We define M theory on $\Rf^9\times S^1\times(S^1/\Zf_2)$ by
periodically identifying the coordinates $x^9$ and $x^{10}$ with
periodicities $2\pi R_9$ and $2\pi R_{10}$, and also identifying
$x^{10}\simeq-x^{10}$ and requiring that the M theory 3-form
$C_{\mu \nu \rho}$ change sign under this reflection. This space
has two boundaries/orientifold planes, at $x^{10}=0$ and at
$x^{10}=\pi R_{10}$. The orientifold breaks half of the eleven
dimensional supersymmetry, leaving 16 supercharges.
%Unlike the smooth case,
This eleven dimensional orientifold is not anomaly-free: there is
a gravitational anomaly in the 10d theory obtained by reducing on
$x^{10}$ that comes from the boundaries, and must be cancelled by
additional massless modes that are restricted to the fixed planes.
As in the 10 dimensional case, part of the anomaly can be
cancelled by a generalized Green-Schwarz mechanism (using the
$C_{\mu\nu(10)}$ form), and the remaining anomaly is cancelled by
the addition of $496$ vector multiplets. In ten dimensions this
restricts the gauge group to be either $\gSO{32}$ or $E_8\times
E_8$. However, in the 11d case the anomaly must be divided equally
between the two fixed planes, so there is \cite{Horava:1995qa} an
$E_8$ gauge group living at each orientifold plane. When the two
radii are large (compared to the 11d Planck length), the low
energy limit is described by 11d supergravity on the cylinder,
coupled to two $\Nc=1$ $E_8$ SYM theories on the two boundaries.

In the limit where $R_9$ is large and $R_{10}$ is small, one obtains \cite{Horava:1995qa} the heterotic
$E_8\times E_8$ string, with string coupling $g_h=R_{10}^{3/2}$ and string length $R_{10}^{-1/2}$ (here and
henceforth we suppress numerical constants of order one, and measure all lengths in 11d Planck units). This is a
valid description of the physics as long as $R_{10}\ll1$ and $R_{9}\gg R_{10}^{-1/2}$ .

When we continue to shrink $R_{10}$ to make it smaller than
$R_9^{-2}$, we reach a point where the $x^9$ circle becomes small
compared to the string scale. At this point we must switch to the
T-dual picture. Recall that the $E_8\times E_8$ heterotic string
with no Wilson lines is T-dual to itself, with an enhanced gauge
symmetry at the self-dual radius. Thus, the appropriate description
in this regime is once again the heterotic $E_8\times E_8$ string,
compactified on a circle of radius $R_{10}^{-1}R_9^{-1}$ with string
coupling $g_{h'}=R_{10}R_9^{-1}$. This description is valid for
$R_{10}\ll R_9$ and $R_9\ll R_{10}^{-1/2}$. In the low-energy
effective action, the $E_8\times E_8\times\gU{1}^2$ gauge group is
enhanced to $E_8\times E_8\times\gSU{2}\times\gU{1}$ along the line
$R_{9}^2R_{10}=1$.

This description is valid for arbitrarily small $R_{10}$, so next we
fix $R_{10}$ and shrink $R_9$. This has the effect of increasing the
string coupling in the T-dual heterotic picture. For $R_9\ll R_{10}$
we open up an extra dimension (as above) and get another region of
moduli space that is also described by M theory on a cylinder. The
length of the dual cylinder is $R_{10}^{1/2}R_9^{-1}$, the radius is
$R_{10}^{-1}R_9^{-1}$ and the Planck length of this ``other M
theory'' is $(l_p)_{M'}=R_9^{-1/3}R_{10}^{-1/6}$. Thus, this
description is valid for $R_9 \ll R_{10}\ll R_9^{-4/5}$.

We are left with the region of $R_{10}\gg R_9^{-4/5}$ and $R_9\ll1$.
This region is covered by the backgrounds that we get by reducing M
theory on the periodic direction of the cylinder. This theory is
known as type I' strings \cite{Dai:1989ua}, and it may be viewed as
an orientifold of type IIA string theory, obtained by dividing by
worldsheet parity together with ${\hat x}^9 \to -{\hat x}^9$. The
fixed points are now orientifold 8-planes ($O8$ planes) of type
$O8^-$, which carry $(-8)$ units of D8-brane charge. Tadpole
cancellation then requires that this background must include also 16
D8-branes.

Another way to obtain the same type I' theory is as the T-dual of
type I string theory. This is helpful in understanding an important
feature of this background \cite{Polchinski:1995df}. In type I
strings, there are two diagrams that contribute to the dilaton
tadpole -- the disk and the projective plane -- and these diagrams
conspire to cancel, homogeneously throughout space-time. On the
other hand, in type I' string theory there are two identical $O8$
planes at the boundaries, and $16$ D8-branes that are free to move
between the boundaries. As a result tadpole cancellation does not
occur locally in this theory: the oriented disk diagram gets a
contribution that is localized at the D8-branes, while the
unoriented projective plane diagram gets contributions localized at
the orientifold planes. Each D-brane cancels one-eighth of the
contribution of an orientifold plane, and there is generally a
gradient for the dilaton, whose exact form depends on the
configuration of D8-branes.

The configuration of the branes between the orientifold planes also
determines the low-energy gauge group of the background. One
possible configuration of D-branes is to have 8 D8-branes on each
$O8$ plane. In this case the dilaton is constant and there is an
$\gSO{16}$ gauge group at each of the boundaries. In all other cases
the dilaton varies between the orientifold planes and the D-branes,
and between the D-branes, with a gradient proportional to the
inverse string length and to the local ten-form charge. This dilaton
gradient causes the dilaton to diverge when the distance between two
such planes is of order $g_{I'}^{-1}l_{s}$ (where $g_{I'}$ is the
string coupling somewhere in the interval), imposing restrictions on
the length of the interval and on the distances between the
orientifold planes and the D-branes.

The last piece of information on type I' string theory that we need
is that when the string coupling becomes infinite on one of the
orientifold planes, there may be D0-branes that become massless
there \cite{Morrison:1996xf,Bergman:1997py,Bachas:1997kn}. If there
are $n$ D8-branes on this orientifold plane, then these additional
light degrees of freedom conspire to enhance the $\gSO{2n}$ gauge
group to $E_{n+1}$. Specifically, in order to get an $E_8$ gauge
group in this theory, we need to put 7 D8-branes on an orientifold
plane, and one D8-brane away from it, precisely at a distance that
will maintain the infinite string coupling at the $O8$ plane. If we
do this at both ends (schematically: ($O8$+7D8)-D8-D8-($O8$+7D8)) we
get an $E_8\times E_8\times\gU1\times\gU1$ gauge group (in nine
dimensions), providing the vacuum of type $I'$ string theory that is
dual to the heterotic $E_8\times E_8$ string with no Wilson line
turned on. The distance in string units between the two single
D8-branes will be denoted by $x_{I'}$.

After this detour on the general properties of type I' string
theory, let us now return to the compactification of M theory to
this background. We can reach a type I' background in two ways:
starting with the original M theory and reducing on $x^9$, or
starting from the dual M theory and reducing on the periodic
direction there. Both constructions give us a type $I'$ theory in
its $E_8\times E_8$ vacuum, with couplings:
\begin{align}\label{I'par}
    &g_{I'1}=R_9^{3/2} && (l_s)_{I'1}=R_9^{-1/2} && R_{I'1}=R_{10}\\
    &g_{I'2}=R_9^{-1}R_{10}^{-5/4} && (l_s)_{I'2}=R_{10}^{1/4} &&
    R_{I'2}=R_{10}^{1/2}R_{9}^{-1}
\end{align}
Note that we need to be careful about what we mean by $g_{I'}$,
since the string coupling varies along the interval and diverges at
the boundaries; what we will mean by $g_{I'}$ (here and in other
cases with a varying dilaton) is the string coupling somewhere in
the interior of the interval, and the differences between different
points in the interval are of higher order in $g_{I'}$ so our
expression is true in the weak $g_{I'}$ limit (which is the only
limit where $g$ is well-defined anyway). Naively, the first
description is valid (except near the orientifold planes where the
string coupling diverges) whenever $R_9 \ll 1$ and $R_{10} \gg
R_9^{-1/2}$, and the second description is valid whenever $R_{10}
\gg R_9^{-4/5}$ and $R_{10} \gg R_9^4$, but this would give an
overlapping range of validity to the two different descriptions
(which would also overlap with some of our previous descriptions).
However, requiring that the distance between each D8 and the $O8$
should be such that the orientifold plane is at infinite coupling,
we find $x_{I'}$ in the two cases to be
\begin{align}
    &1) \qquad x_{I'1}=(R_{10}R_9^2-1)R_9^{-3/2}\\
    &2) \qquad x_{I'2}=(1-R_9^2R_{10})R_{9}^{-1}R_{10}^{1/4}
\end{align}
Now, the regime of validity for each of the type $I'$ backgrounds
(defined by $(g_s)_{I'}\ll1$ and $x_{I'}\geq 0$) is distinct; the
first description is valid when $R_9 \ll 1$ and $R_{10} \leq
R_9^{-2}$ and the second description is valid when $R_{10} \gg
R_9^{-4/5}$ and $R_{10} \geq R_9^{-2}$. Furthermore, at the line
$R_{10}R_9^2=1$, which was the line of enhanced symmetry in the
heterotic $E_8\times E_8$ string theory, we have $x_{I'}=0$ for both
backgrounds. On this line the two D8-branes coincide, enhancing the
$\gU1\times\gU1$ symmetry to $\gSU{2}\times\gU1$ exactly as in the
heterotic case. This implies that the two type $I'$ backgrounds are
actually related by an $\gSU{2}$ gauge transformation, which is
consistent with the fact that $x_{I'1}/(l_s)_{I'1} =
-x_{I'2}/(l_s)_{I'2}$. We see that the line $R_{10} R_9^2 = 1$
continues to be a line of enhanced symmetry throughout the moduli
space.

The backgrounds described thus far cover the whole moduli space of
nine dimensional $\Nc=1$ backgrounds with an $E_8\times E_8$ gauge
group. The top graph in figure \ref{modulifig16} displays how
these backgrounds fill in all possible values of the radii of the
M theory we started with. Of course, the top part of this figure
(above the dashed line) is related by an $\gSU2$ gauge
transformation to the bottom part, so the true moduli space is
just half of this figure (above or below the dashed line).

\FIGURE[hbt]{
\includegraphics[height=3in]{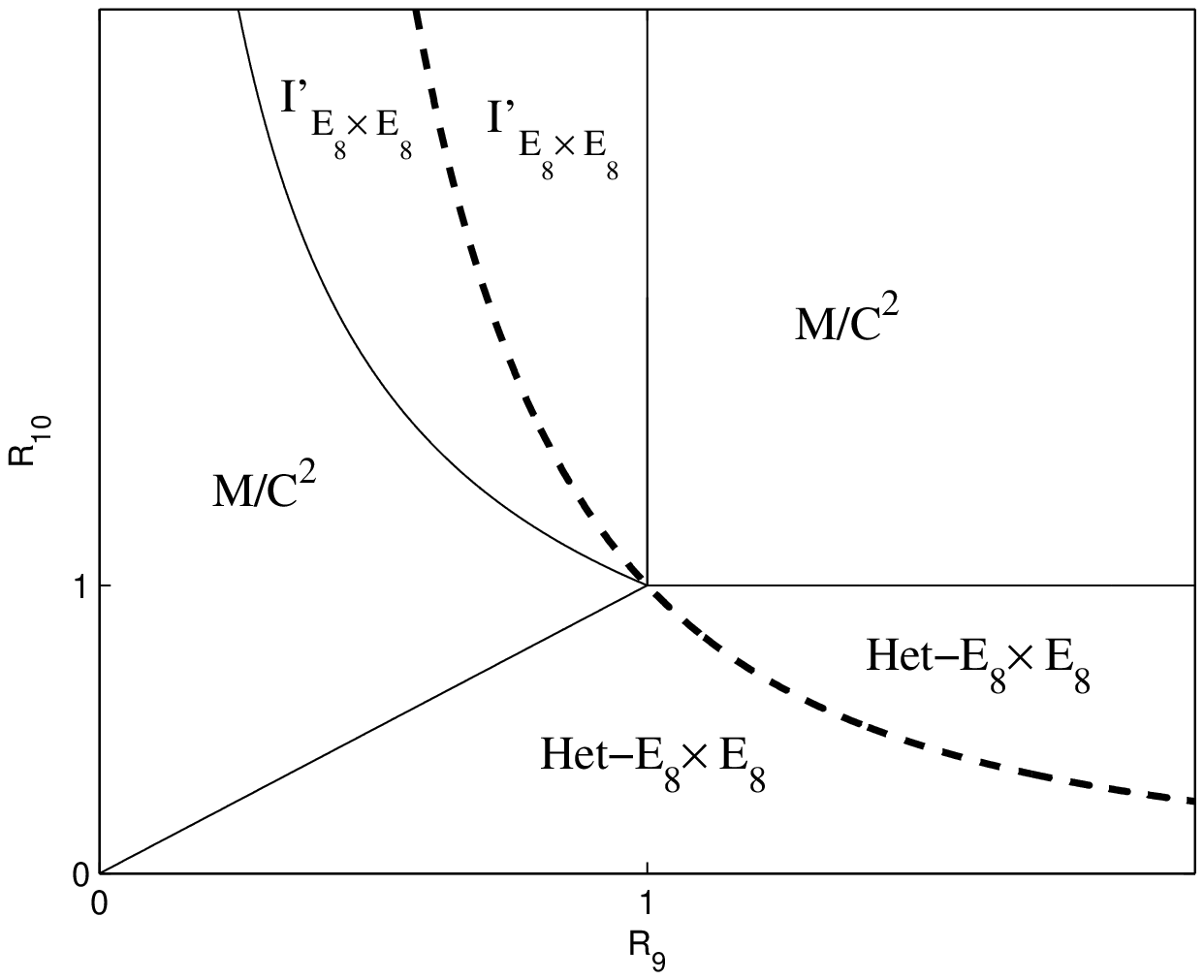}
\includegraphics[height=3in]{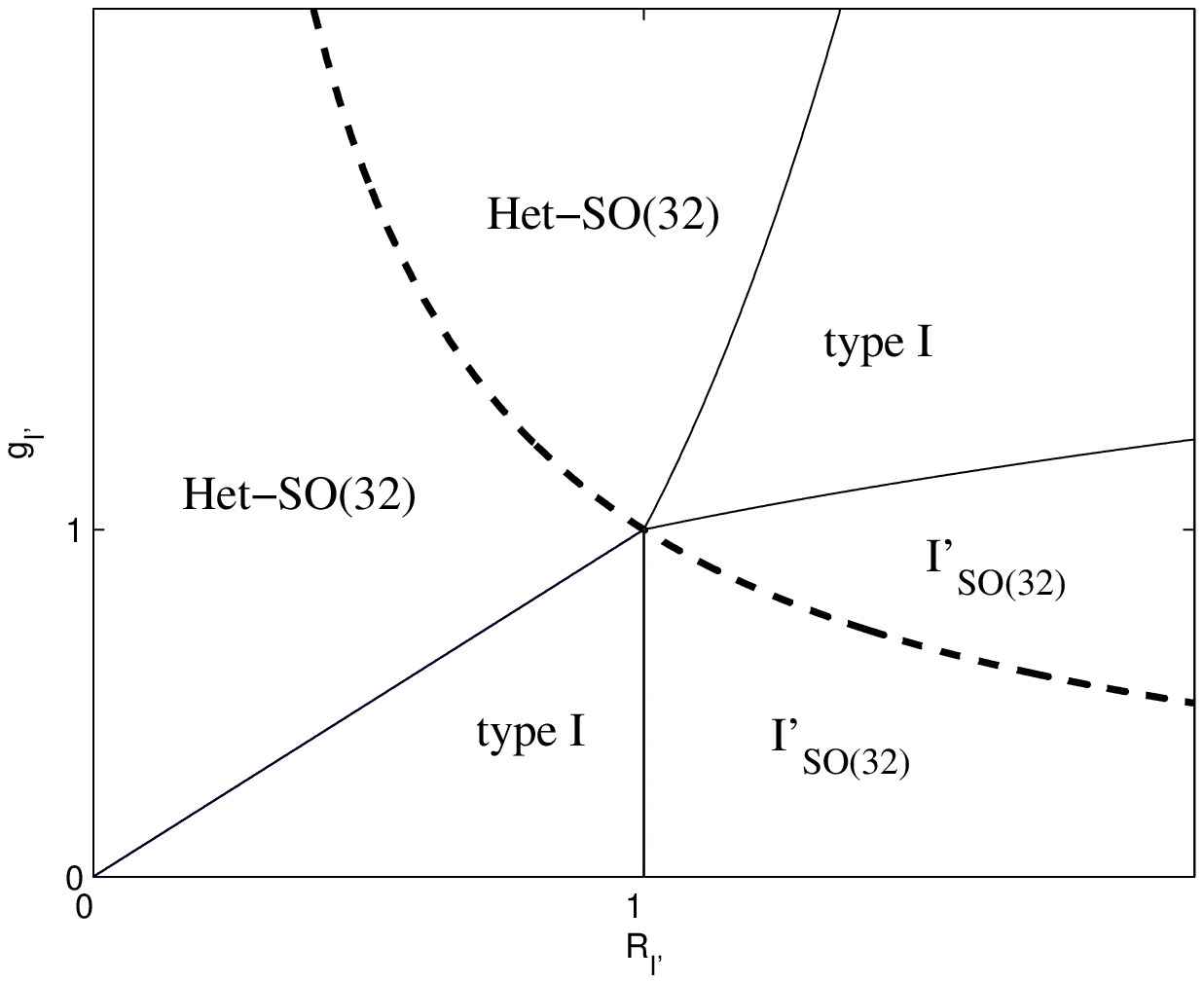}
\caption{The $E_8\times E_8$ and $\gSO{32}$ subspaces of the
moduli space of rank $18$ compactifications of M theory and string
theory. In the first graph the parameters are the period and
length of the cylinder in Planck units as defined for the upper
right compactification of M theory on a cylinder, and in the
second they are the distance between the orientifold planes in
units of the string length, and the string coupling on the $O8^-$
plane with the branes on it, of the type I' background appearing
in the lower right corner. The dashed line in both graphs is the
line of enhanced $\gSU2$ symmetry; the regions of the graphs below
and above this line are identified.} \label{modulifig16} }

Next, we turn to the subspace of the rank $18$ backgrounds with an
unbroken $\gSO{32}$ gauge symmetry.\footnote{Actually we will only
describe one connected component of this subspace, which will turn
out to be analogous to the rank $2$ case. The other component has a
discrete Wilson line that is only felt by spinor representations of
$\gSO{32}$.} This subspace includes the type $I'$ background in its
$\gSO{32}$ vacuum. This vacuum is obtained by having all $16$
D8-branes sit at one of the $O8^-$ planes (so that it has the same
charges as an $O8^+$ plane). The dilaton then grows as we go from
this $O8^-$ plane to the other one. We denote the distance between
the orientifold planes as $R_{I'}$ (in string units) and the string
coupling (defined, for instance, near the $O8^-$ plane with the
branes) as $g_{I'}$. The regime of validity of this description is
$1\ll R_{I'}\ll g_{I'}^{-1}$.

For $R_{I'}\ll 1$, we need to switch to the T-dual of this
description, which is the type I background with no Wilson lines.
The string coupling in type I is $g_{I}=g_{I'}/R_{I'}$ and the
radius of the compact circle is $R_I=R_{I'}^{-1}$. This
description is thus valid for $g_{I'}\ll R_{I'}\ll1$.

As we further decrease $R_{I'}$ we increase the coupling of the type I string theory, and eventually we should
go over to the S-dual heterotic $\gSO{32}$ string. The parameters of this heterotic background are
$g_h=R_{I'}/g_{I'}$ and $R_h = 1 / R_{I'}$, and its string length is $(l_s)_h = (g_{I'}/R_{I'})^{1/2}$, implying
that it is a valid description for $R_{I'}\ll g_{I'}\ll R_{I'}^{-1}$.

On the line $g_{I'}= R_{I'}^{-1}$ we find $R_h = (l_s)_h$. This is a
line of self T-duality and enhanced $\gSU2$ gauge symmetry in the
heterotic $\gSO{32}$ string. We can now move to the T-dual picture,
obtaining another heterotic $\gSO{32}$ description. By going to the
strong coupling limit of the new heterotic background, we get
another type I background, and by another T-duality we get another
regime described by the type $I'$ background. These three types of
backgrounds cover the whole $\gSO{32}$ subspace of the moduli space,
as presented in the second graph of figure \ref{modulifig16}.

In the heterotic $\gSO{32}$ description we found that along the line
$R_{I'}=g_{I'}^{-1}$ a $\gU{1}$ symmetry gets enhanced to $\gSU2$,
by the usual mechanism of a winding mode becoming massless at the
self-dual point. As in the previous example, the same enhancement
occurs on this line also in other regions of the moduli space. In
the type $I'$ description, this is the line where the $O8$ plane
without the branes is at infinite coupling. The half-D0-branes that
live on the orientifold plane then become massless at this line, and
they are responsible for the symmetry enhancement (from $U(1)$ to
$E_1=\gSU{2}$) in this description.

It is important to remember that the two branches depicted in figure
\ref{modulifig16} are just specific subspaces of the moduli space of
rank 18 theories and that the rest can be reached by turning on
Wilson lines in the heterotic or type I pictures, or equivalently
moving around D8-branes in the type $I'$ picture. Thus, the two
plots in the figure are just two slices of the full moduli space.
This is important to emphasize since in the next subsection a
similar picture will arise, but in that case there are no Wilson
lines to be turned on, so the two branches are actually two
disconnected components of the 9d moduli space.

%%%%%%%%%%%%%%%%%%%%
%
%
%%%%%%%%%%%%%%%%%%%%
\subsection{M theory on a Klein bottle and other rank $2$ compactifications}\label{monk2subsec}

We now wish to repeat the above analysis for M theory compactified on a Klein bottle (first considered in
\cite{Dabholkar:1996pc}) instead of a cylinder. One may start by asking if M theory makes sense at all on a
Klein bottle. We will see that the answer is yes, and that this background arises as a strong coupling limit of
consistent string theory backgrounds.

Let us begin by considering type IIB string theory gauged by the
symmetry group $\Zf_2=\{1,H_9\Omega\}$, where $\Omega$ is
worldsheet parity and $H_9$ is a shift in the 9th coordinate,
\begin{equation}
    H_9\Omega\,:\,
    \begin{cases}
    X^\mu(z,\bar{z})\simeq X^\mu(\bar{z},z), & \mu\neq9\\
    X^9(z,\bar{z})\simeq X^9(\bar{z},z)+2\pi R_9.
    \end{cases}
\end{equation}
This orientifold theory is derived by imposing $H_9\Omega=1$ on the spectrum of the IIB theory compactified on a
circle of radius $2R_9$; this breaks half of the supersymmetry (one of the gravitinos is projected out and the
other one remains). In this case there are no fixed planes, and one does not need to add D-branes to cancel any
tadpole, as explained in \cite{Dabholkar:1996pc}. We refer to this background as the Dabholkar-Park background
or the DP background.

We can compactify the DP background on an additional circle of
radius $\tilde{R}_8$ in the $x^8$ direction, and T-dualize in this
direction. Because of the orientation reversal in the original DP
background, the resulting background is type IIA string theory
gauged by
\begin{equation}\label{our IIA}
%    H_9\Omega\,:\,
    \begin{cases}
    X^\mu(z,\bar{z})\simeq X^\mu(\bar{z},z), & \mu\neq8,9\\
    X'^8(z,\bar{z})\simeq -X'^8(\bar{z},z),\\
    X^9(z,\bar{z})\simeq X^9(\bar{z},z)+2\pi R_9.
    \end{cases}
\end{equation}
Thus, this describes a type IIA compactification on a Klein bottle
($K2$) of area $(2\pi R_8)\times(2\pi R_9)$, where $R_8 = l_s^2 /
\tilde{R}_8$, with an orientation reversal operation (see also
\cite{Keurentjes:2000bs}).\footnote{ The orientation reversal leads
to various periodicity conditions for the $p$-form fields, that are
explained in detail in appendix \ref{pformrules}.} One can lift this
theory to a background of M theory in the usual way by taking the
strong coupling limit (the supersymmetry of this background
prohibits a potential for the dilaton). The worldsheet parity is
identified in M theory with flipping the sign of all components of
the 3-form field. Thus, as promised, M theory on a Klein bottle
naturally arises as a strong coupling limit of a string theory.

Consider now M theory compactified on a Klein bottle of radii
$K2(R_{10},R_9)$ measured in Planck units, namely with the
identification
\begin{equation}\label{k2defr}
    (X^{10},X^9)\simeq(-X^{10},X^9+2\pi R_9),
\end{equation}
including a reversal of the $3$-form field, and also with $X^{10}$
periodically identified with radius $R_{10}$. If we shrink
$R_{10}$ we get a background that is described in detail in
\cite{Gutperle:2000bf} and \cite{Hellerman:2005ja} (additional
related analysis can be found in \cite{Vafa:1995gm}). This
background can be defined as type IIA string theory with a gauging
of the symmetry
\begin{equation}\label{defAOA}
    (-)^{F_L}\times (X^9\rightarrow X^9+2\pi R_9),
\end{equation}
where $F_L$ is the space-time fermion number of the left-moving
fields on the worldsheet. We refer to this background as the
asymmetric orbifold of type IIA or AOA. The type IIA string
coupling is $g_s \simeq R_{10}^{3/2}$, and $l_s \simeq
R_{10}^{-1/2}$.

If we continue to shrink $R_{10}$, the circle becomes small (in
string units) and we need to change to the T-dual description. It
was noticed in \cite{Gutperle:2000bf} that the AOA background has
an enhanced $\gSU2$ symmetry point when $R_9=l_s/\sqrt2$, and in
\cite{Hellerman:2005ja} it was demonstrated that (as implied by
this enhanced symmetry) this background is actually self-T-dual.
In appendix \ref{mod and T in partition} we explicitly evaluate
the partition function of this model, showing that it is modular
invariant and respects the T-duality. Therefore, when we continue
to shrink $R_{10}$ we should switch to a T-dual AOA description,
which has $(R_9)_T \simeq 1/R_9 R_{10}$ and $(g_s)_T \simeq
R_{10}/R_9$. This dual AOA description can in turn be lifted to a
dual M theory on a Klein bottle. The regimes of validity of these
different descriptions are exactly the same (up to numerical
constants of order one) as those we found in the previous
subsection, with M theory on the Klein bottle replacing M theory
on a cylinder, and the AOA background replacing the heterotic
$E_8\times E_8$ string.

Going back to M theory on a Klein bottle, what happens if we shrink
$R_9$? Naively, we get a theory that is an orientifold of IIA on an
$S^1/\Zf_2$, with some boundary conditions at the fixed planes.
However, we have a $\Zf_2$ symmetry that exchanges the two
boundaries, and since there cannot be any tadpole for the RR
10-form, we can only have neutral orientifold planes (with respect
to the 10-form charge) at the boundaries, unlike the standard $O8^-$
planes (which carry $(-8)$ units of charge) and $O8^+$ planes (which
carry $+8$ units of charge).\footnote{Notice that this contradicts a
suggestion occasionally found in the literature, that in this limit
we get a background defined by IIA on an $S^1/\Zf_2$ with an $O8^-$
plane on one end and an $O8^+$ plane on the other.} For now we
conjecture that the reduction in this direction leads to a
background which we call the $X$-background, and that the line of
enhanced symmetry $R_{10}R_9^2=1$ is also a line of enhanced
symmetry in the $X$-background. Thus, the two remaining regions of
the moduli space are covered by the $X$-background and its dual
(obtained by reducing the dual M theory on the periodic cycle). In
\S\ref{thexbg} we shall propose a stringy description of this
background.

The theories described above (starting with M theory on a Klein
bottle) cover a complete moduli space of the form
$\gSO{1,1,\Zf}\backslash\gSO{1,1}\times \Rf$, as shown on the top
graph of figure \ref{modulifigk2}. However, there are several
additional string backgrounds with 16 supercharges and a rank 2
gauge symmetry that were not included so far, so they must be in a
separate component of the moduli space. The first is the DP
background described above. Next, there is the orientifold of type
IIA on an interval $S^1/\Zf_2$, with an $O8^-$ plane on one end and
an $O8^+$ plane on the other \cite{Witten:1997bs}. We refer to this
background as the $O8^{\pm}$ background, and denote the length of
the interval in string units by $\pi R_{\pm}$. No D-branes are
required for tadpole cancellation here. However, the dilaton does
have a gradient (identical to that in the $\gSO{32}$ background of
the type $I'$ string) that puts a limit on the maximum length of the
interval, of the form $R_\pm<1/g_\pm$ (where $g_\pm$ is again
defined as the string coupling somewhere in the interval). This
background was studied in \cite{Witten:1997bs}, where it was shown
to be T-dual to the DP background.  An important feature of this
background is that when the distance between the orientifold planes
is such that the coupling on the $O8^-$ is infinite, there are
half-D0-branes stuck on the orientifold that become massless.
Locally, this is exactly the same as in the type $I'$ $\gSO{32}$
vacuum; in both cases the fractional branes enhance the $\gU{1}^2$
symmetry to $\gSU2\times\gU1$.

\FIGURE[hbt]{
\includegraphics[height=3in]{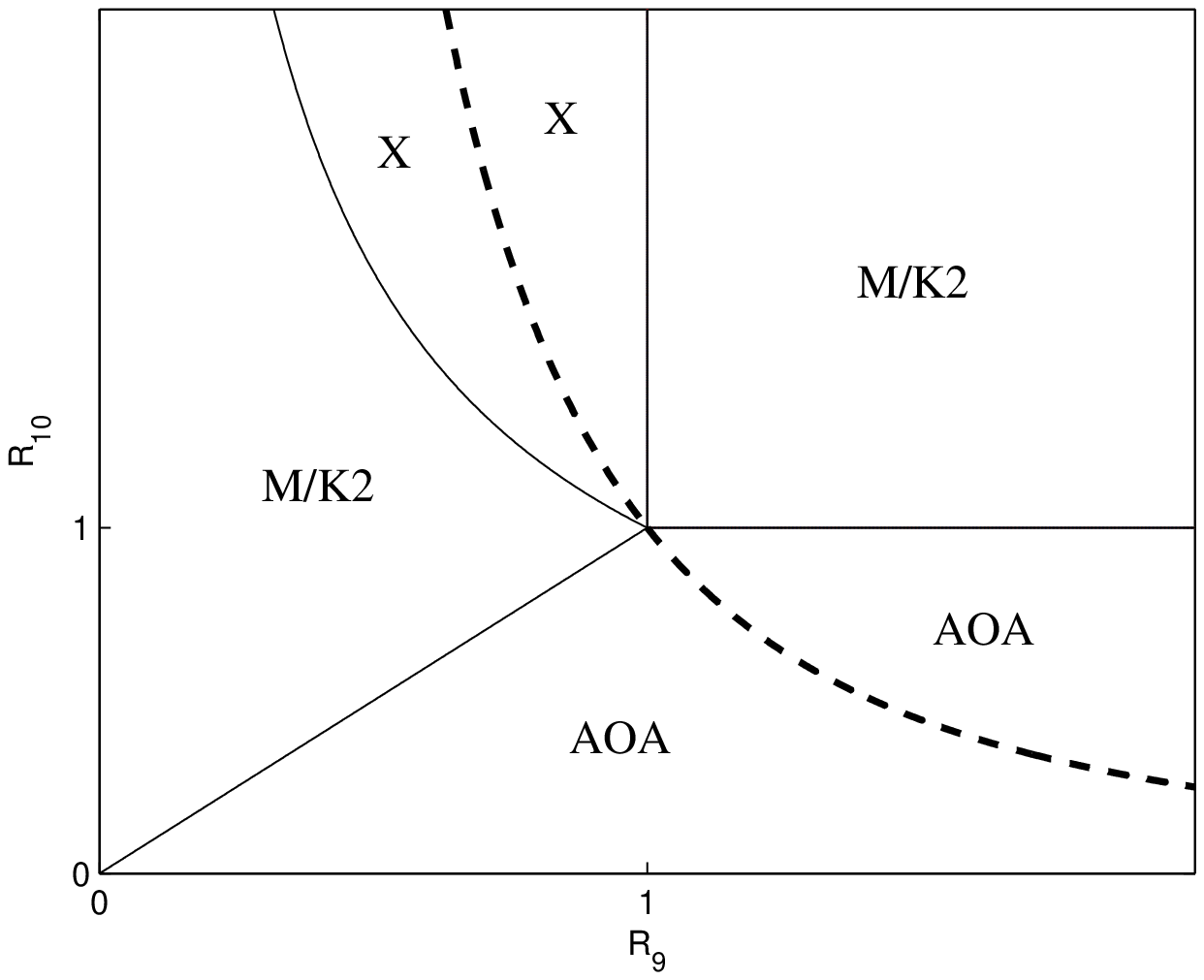}
\includegraphics[height=3in]{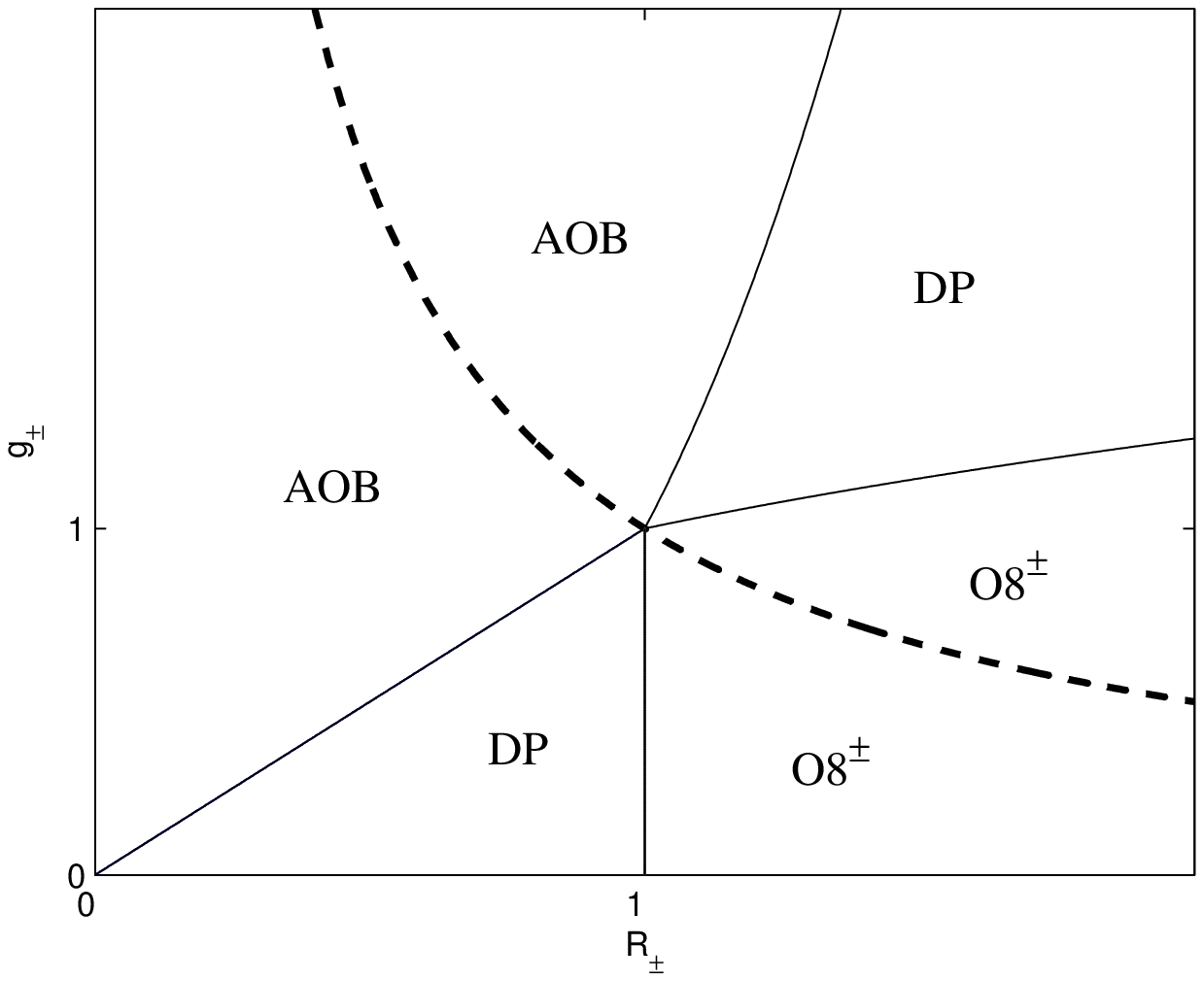}
\caption{The two disconnected components of the moduli space
including the backgrounds of subsection \ref{monk2subsec}. In the
first graph the parameters are the periods of the Klein bottle in
Planck units as defined in \eqref{k2defr} for the upper right M
theory description. In the second graph, the parameters are the
radius of the circle in string units and the string coupling (in
the middle of the interval) for the $O8^{\pm}$ background on the
lower right corner. The dashed line in both graphs is the line of
enhanced $\gSU2$ gauge symmetry, and the backgrounds below the
line are identified with the backgrounds above the
line.}\label{modulifigk2}}

Finally, one other background can be obtained by gauging type IIB
string theory by the symmetry \eqref{defAOA}. We call this
background the Asymmetric Orbifold of type IIB or AOB. Recalling
the following property of IIB strings
\begin{equation}
    (-1)^{F_L}=S\Omega S
\end{equation}
(where $S$ is the S-duality transformation of type IIB string
theory) and applying the adiabatic argument of \cite{Vafa:1995gm},
we see that this background is S-dual to the DP background.
Exactly like the AOA background, the AOB background is self-T-dual
\cite{Hellerman:2005ja} and has an enhanced $\gSU{2}$ symmetry
when the radius of the circle is $l_s/\sqrt2$.

We can now describe the second component of the moduli space. Start
with the $O8^\pm$ background with string coupling $g_\pm$ and radius
$R_\pm$ (measured in string units). This is a valid description for
$g_\pm\ll1/R_\pm$ and $R_\pm\gg1$. As we decrease $R_\pm$, we need
to T-dualize, as in \cite{Witten:1997bs}, and we get the DP
background with radius\footnote{We continue to ignore numerical
constants of order one.} $1/R_\pm$ and string coupling
$g_\pm/R_\pm$. The DP description is valid as long as $g_\pm\ll
R_\pm\ll1$. We can further decrease $R_\pm$ to the point where the
DP string coupling is too large. At this point we turn to the S-dual
picture, which is the asymmetric orbifold of type IIB (AOB). The
string coupling is now $R_\pm/g_\pm$, and the S-dual string length
is $(g_\pm/R_\pm)^{1/2}$, such that the radius $1/R_\pm$ in string
units is now $(g_\pm R_\pm)^{-1/2}$. This is a good description for
$g_\pm\ll1/R_\pm$ and $R_\pm\ll g_\pm$.

We continue next by increasing $g_\pm$ to the point where the AOB
radius becomes small. Then we must use the self T-duality of this
background to arrive at a dual AOB description, with radius
$(g_\pm R_\pm)^{1/2}$ (in units of its string length) and string
coupling $g_\pm^{-1/2}R_\pm^{3/2}$. This description is good for
$g_\pm\gg 1/R_\pm$ and $g_\pm\gg R_\pm^3$. The next step is to
take the S-dual of the dual AOB background. This gives another DP
background with radius (in units of its string length)
$g_\pm^{3/4}R_\pm^{-1/4}$ and coupling $g_\pm^{1/2}R_\pm^{-3/2}$.
This description is valid for $g_\pm\gg R_\pm^{1/3}$ and $g_\pm\ll
R_\pm^{3}$. The last step is to T-dualize the new DP background to
a dual $O8^{\pm}$ background, with radius
$g_\pm^{-3/4}R_\pm^{1/4}$ (in units of its string length) and
coupling $g_\pm^{-1/4}R_\pm^{-5/4}$. This description is valid for
$g_\pm\ll R_\pm^{1/3}$ and $g_\pm\gg1/R_\pm$.

These six string theories cover the entire range of values of
$(g_\pm,R_\pm)$. The reader should notice that the line of enhanced
$\gSU2$ symmetry in the $O8^{\pm}$ background, $R_\pm \sim 1/g_\pm$,
smoothly goes into the line of enhanced $\gSU2$ symmetry of the AOB
background \cite{Gutperle:2000bf}, as in the previous cases we
discussed.

Figure \ref{modulifigk2} summarizes the structure of the moduli space. As promised, it has two disconnected
components in the nine dimensional sense. Note that the nine dimensional low-energy effective action on the two
components is identical, but the massive spectrum is different. The components can be related by compactifying
on an additional circle and performing a T-duality, but they are not connected as nine dimensional backgrounds.
There is an obvious relation between each of these components and one of the subspaces of the moduli space of
rank 18 compactifications discussed in the previous subsection and depicted in figure \ref{modulifig16}.
%%%%%%%%%%%%%%%%%%%%%%%%%
%
%
%%%%%%%%%%%%%%%%%%%%%%%%%
\subsection{The $X$ background demystified}\label{thexbg}

In the previous subsection we left open the description of the limit of the Klein bottle where $R_9$ is small
and $R_{10}$ is large. In this limit the Klein bottle geometrically looks as in figure \ref{Kleinbottle} (though
the geometric description is not really valid at distances smaller than the 11d Planck scale). This limit should
correspond to some $10$ dimensional string theory; let us collect some features of this theory:
\FIGURE[h]{
\includegraphics[height=2cm]{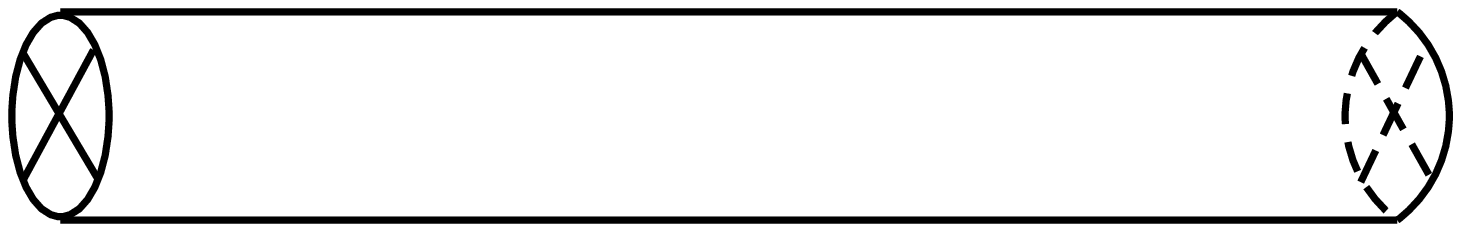}
\caption{In the $X$ region, the Klein bottle looks like a long tube
between two cross-caps.} \label{Kleinbottle}}
\begin{itemize}
\item After reducing M theory on the small circle $R_{9}$, we
expect to obtain a type IIA string theory that lives on
$\mathbb{R}^{8,1}\times I$ where $I$ is an interval. On each
boundary of the interval we should have an orientifold plane;
however these orientifold planes are non-standard because the
orientifolding is accompanied by a half-shift on the M theory
circle. This half-shift modifies, among other things, the
properties of D0-branes near the orientifold. One can think of
this shift as a discrete RR flux characterizing the orientifold
plane, which changes its charge from the usual charge of $(-8)$ to
zero \cite{Keurentjes:2001cp}. We will denote such orientifold
planes by $O8^0$.

\item The fact that the orientifold carries no D8-brane charge is
consistent with tadpole cancellation of the 10-form field and the
dilaton. Note that there is a $\mathbb{Z}_2$ symmetry exchanging
the two ends of the interval, so the two orientifold planes must
carry the same charge (unlike the case in the $O8^{\pm}$
background described in the previous subsection).

\item Naively one expects that in the limit of small $R_9$ the 10
dimensional string theory can be made very weakly coupled, and the
description should involve the usual type IIA string theory, at
least far from the boundaries of the interval (where there may or
may not be large quantum corrections). We will see that there are
some regions of the moduli space where this naive expectation
fails.

\item The discussion of the previous subsection suggests that we
should have an enhanced $\gSU2$ gauge symmetry when the interval
is of size $R\sim 1/g_s$ (in type IIA string units). One may
expect this enhanced symmetry to come from half-D0-branes on the
orientifold planes, as in some of the examples discussed in the
previous subsections. However, since we have a $\mathbb{Z}_2$
symmetry relating the two orientifolds, it is hard to imagine how
the $\gU{1}^2$ group would be enhanced to $\gU1\times\gSU2$ rather
than to the more symmetric $\gSU{2}^2$.
\end{itemize}

The last item above suggests some modification of the naive
picture of this background. The picture that we will suggest for
the correct description of this background is based on two facts :
\begin{itemize}
\item In our analogy between the rank 18 and the rank 2 theories,
the $X$ background plays the same role as the type $I'$ $E_8\times E_8$ background. In this background the two
orientifold planes are always at infinite coupling, and the enhanced $\gSU2$ symmetry arises when two D8-branes
in the middle of the interval come together \cite{Bachas:1997kn}.

\item In some cases, when a standard ($O8^-$) orientifold plane of
charge $(-8)$ is at infinite coupling, it can emit a D8-brane into
the bulk, leaving behind an orientifold plane of charge $(-9)$ which
has no perturbative description (and which always sits at infinite
coupling). This phenomenon cannot be seen in perturbation theory,
but it can be deduced from an analysis of D4-brane probes
\cite{Morrison:1996xf}.
\end{itemize}

Our suggestion is that each $O8^0$ plane non-perturbatively emits a
D8-brane and becomes an $O8^{(-1)}$ plane which always sits at
infinite coupling. The moduli space and 10-form fluxes of this
system are then identical to those of the $E_8\times E_8$ type $I'$
background, with the $O8^{(-1)}$ plane playing the same role as the
$O8^-$ plane with $7$ D8-branes on it. Now, the gauge symmetry is
enhanced to $\gSU2$ in a $\mathbb{Z}_2$-symmetric manner, when the
D8-branes meet in the middle of the interval, and this enhancement
is perturbative (it can happen at weak coupling). Denoting the
string coupling in the region between the two emitted D8-branes by
$g_s$ (it does not vary in the interval between the D8-branes), the
distance of each brane from the respective $O8^{(-1)}$ plane is
$\sim 1/g_s$. Thus, when the branes meet, our interval indeed has a
size proportional to $1/g_s$ as required.

We suggest that this is the correct description of $O8^0$
orientifold planes. Note that such a large non-perturbative
correction to the description of high-dimensional orientifold planes
is not surprising; already in the case of $O7$ planes it is known
\cite{Sen:1996vd} that they non-perturbatively split into two
7-branes, and the corrections to $O8$ planes are expected to be even
larger. Our suggestion implies non-trivial corrections to
compactifications of M theory involving crosscaps (similar
corrections in M theory should also occur in a compactification on a
M\"obius strip, as will be discussed in the next subsection). These
corrections are similar to the ones that occur for M theory on a
cylinder with no Wilson lines. Note that when we are close to the
enhanced $\gSU2$ point, the corrections to the naive M theory
picture shown in figure \ref{Kleinbottle} are not just localized
near the cross-caps as one may naively expect, but the string
coupling actually varies along the whole interval.

In order for this proposal to be consistent, there should be no
massless fractional D0-branes on the $O8^{(-1)}$ planes, which
would lead to more enhanced symmetries than we need. Because of
the shift in the M theory circle involved in the orientifolding,
the radius of the cross-cap is actually half of the radius of the
M theory circle in the bulk, which implies that only even momentum
modes (in units of the minimal momentum on a standard orientifold
plane) are allowed there. Hence, there are no half D0-branes in
backgrounds with $O8^0$ orientifolds (or $O8^{(-1)}$
orientifolds).
%%%%%%%%%%%%%%%%%%%%%%%%%
%
%
%%%%%%%%%%%%%%%%%%%%%%%%%
\subsection{M theory on a M\"obius strip and other rank $10$ compactifications}

There is one additional disconnected component of the moduli space of nine dimensional compactifications with
$\Nc=1$ supersymmetry. Consider the heterotic $E_8\times E_8$ string compactified to nine dimensions on a circle
of radius $R_9$. One can consider an orbifold of this theory generated by switching the two gauge groups
together with a half-period-shift of $x^9$. This theory is known as the CHL string; as usual one should add
twisted sectors for the consistency of the orbifold (for more details see
\cite{Chaudhuri:1995fk,Chaudhuri:1995bf}). This leads to a nine dimensional compactification with a rank 10
gauge group. We will focus on the subspace of the moduli space of this theory in which the $E_8$ symmetry is
unbroken.

An orbifold of the $E_8\times E_8$ heterotic string on a circle
can also be viewed as an orbifold of M theory compactified on a
cylinder. Begin with M theory compactified on the cylinder
\begin{equation}
 x^9\simeq x^9+2\pi R_9 \hspace{1em},\hspace{1em} x^{10}\in
    \left[-\frac{\pi}{2} R_{10},\frac{\pi}{2} R_{10}\right].
\end{equation}
The action on the heterotic string which we described above lifts
in M theory to
\begin{equation}\label{mobdefr}
    x^9\simeq x^9+\pi R_9 \hspace{1em},\hspace{1em}
    x^{10}\simeq -x^{10}.
\end{equation}
Upon identifying points related by this transformation we obtain a
M\"obius strip, with a cross-cap at $x^{10}=0$ and a boundary at
$x^{10}=\pi R_{10}/2$, as depicted in figure \ref{CHLlimit}. Thus,
this component of the moduli space is generated by various limits of
M theory compactified on the M\"obius strip
\cite{Dabholkar:1996pc,Park:1996it}. Notice that anomaly
cancellation as in \cite{Horava:1995qa} tells us that the single
boundary of the M\"obius strip carries an $E_8$ gauge symmetry,
which is consistent with the low-energy gauge symmetry of the CHL
string.

\FIGURE{
\includegraphics[height=2.5cm]{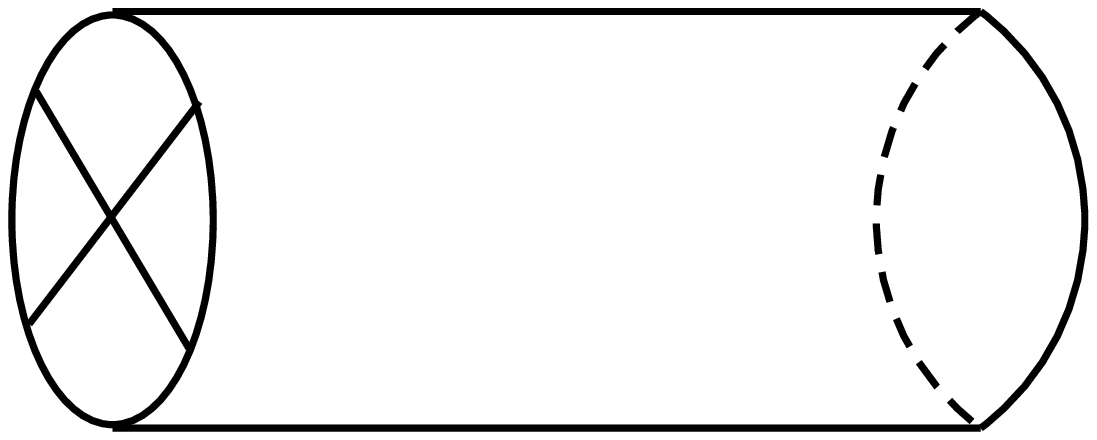}\label{CHLlimit}\caption{M
theory on a M\"obius strip describes the strong coupling limit of
the CHL string.}}

So far we have good descriptions of the regions of moduli space
where both $R_9$ and $R_{10}$ are large, and when $R_{10}$ is small
(leading to the CHL string). It is natural to ask what string theory
backgrounds are obtained when we reduce on the other direction,
$R_{10}\gg l_P \gg R_9$. In this limit, the cross-cap and the
boundary (which is topologically $\mathbb{R}^{8,1}\times S^1$) of
figure \ref{CHLlimit} are very far from each other. The boundary
becomes a usual $O8^{-}$ plane of IIA string theory. Our previous
description of $E_8$ symmetries in type IIA string theory implies
that this $O8^-$ plane has 7 D8-branes on top of it, and there is an
additional one displaced such that the system $O8^{-}+$7D8 is at
infinite string coupling. The cross-cap should behave exactly as in
the case of the Klein bottle, described in the previous subsection.
Therefore, the naive $O8^0$ plane emits an extra D8-brane and
becomes an $O8^{(-1)}$ plane at infinite string coupling.

When we approach the point $R_{10}R_{9}^2\sim 1$, we find that the two D8-branes in the bulk come together, and
enhance the $\gU1\times\gU1$ symmetry to $\gSU2\times\gU1$. As in our previous examples, the same symmetry
enhancement arises also for small $R_{10}$, where it arises at the self-dual radius of the perturbative CHL
string (for an exhaustive analysis of the momentum lattices in toroidal compactifications of CHL strings see
\cite{Mikhailov:1998si}). In fact, it is just the same as the $\gSU2$ enhanced symmetry of the $E_8\times E_8$
string at the self dual radius (the gauge bosons are BPS and survive the CHL projection). This is another
consistency check on our proposal for the behavior of the cross-cap in M theory.

\FIGURE{
\includegraphics[height=3in]{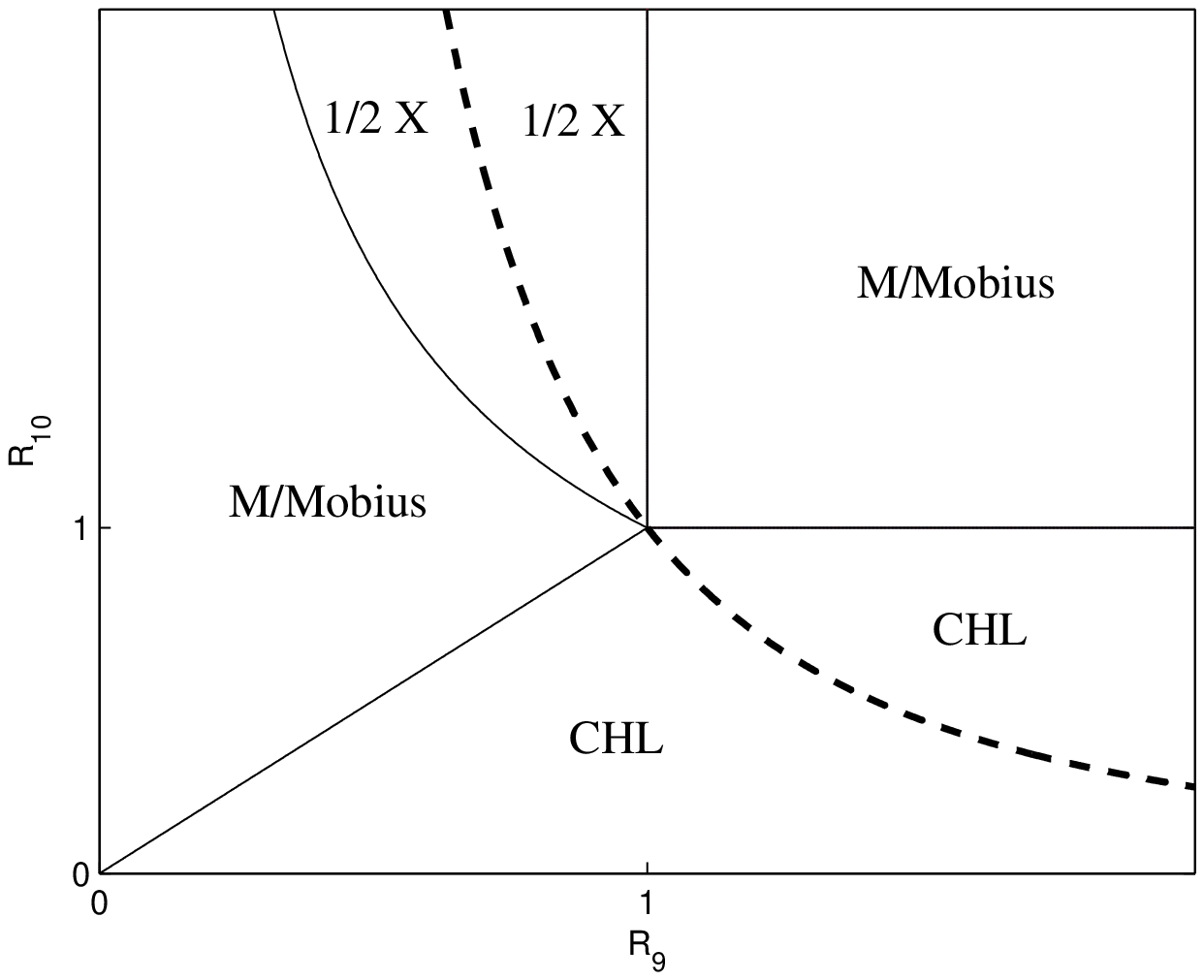}
\caption{The moduli space of compactifications of M theory on a
M\"obius strip. The parameters are the period and length of the
strip in Planck units as defined for the upper right M theory. The
dashed line is the line of enhanced $\gSU2$
symmetry.}\label{modulifigCHL}}

We conclude that the moduli space of M theory on a M\"obius strip
has a line of enhanced $\gSU2$ symmetry, and that all of its limits
may be understood (with some strongly coupled physics occurring in
the IIA limit). This moduli space is depicted in figure
\ref{modulifigCHL}, where the type IIA orientifold limit is denoted
by $(1/2)X$. In fact, this moduli space is identical in structure to
the other moduli spaces we encountered in our survey, arising from M
theory on a cylinder and on a Klein bottle.

%%%%%%%%%%%%%%%%%%%%%%%%%%%%%%%%%%
%
%
%
%%%%%%%%%%%%%%%%%%%%%%%%%%%%%%%%%%
\section{The M(atrix) theory description of M theory on a cylinder
and on a Klein bottle}\label{Ms}

In this section we describe the M(atrix) theory
\cite{Banks:1996vh} which is the discrete light-cone quantization
(DLCQ) of some of the theories described in the previous section
-- M theory compactified on a cylinder and on a Klein bottle. We
begin by considering the case of a cylinder \cite{Kabat:1997za},
and then move on to the Klein bottle. We review in detail the case
of the cylinder because of the great similarity between these two
compactifications (as described in the previous section), which is
useful in the construction of the correct M(atrix) theory of the
Klein bottle.

\subsection{The M(atrix) theory of M theory on a
cylinder}\label{Matcyl}

This subsection is based on \cite{Kabat:1997za}, with the addition
of a systematic derivation of their M(atrix) theory and of some
small corrections to the identifications of parameters presented
in that paper.

M(atrix) theory is the discrete light-cone quantization of M
theory backgrounds \cite{Susskind:1997cw}; it provides the
Hamiltonian for these theories compactified on a light-like
circle, with $N$ units of momentum around the circle. In general,
such a DLCQ description is very complicated. However, in some M
theory backgrounds it simplifies, because a light-like circle may
be viewed \cite{Seiberg:1997ad,Sen:1997we} as a limit of a very
small space-like circle, and M theory on a very small space-like
circle is often very weakly coupled. This leads to a simple
description of the DLCQ Hamiltonian, in which most of the degrees
of freedom of M theory decouple. In particular, this is the case
for the M(atrix) theory of M theory itself, which is given just by
a maximally supersymmetric $\gU{N}$ quantum mechanical gauge
theory, and for the M(atrix) theory of M theory compactified on a
two-torus, which is given by the maximally supersymmetric $\gU{N}$
$2+1$ dimensional gauge theory, compactified on a dual torus.

The generic simplicity of M(atrix) theory is based on the fact that M theory compactified on a small space-like
circle becomes a weakly coupled type IIA string theory. However, when boundaries are present in the M theory
compactification, they usually destroy this simplicity. For instance, as is evident from figure
\ref{modulifig16}, if we take M theory on $S^1/\Zf_2$ and compactify it further on a very small space-like
circle, we do not obtain a weakly coupled background (but, rather, we obtain M theory on a dual cylinder). Thus,
generically the DLCQ of M theory backgrounds with boundaries is very complicated. However, there is an extra
degree of freedom one can use in the DLCQ constructions, which is a Wilson line along the light-like circle;
such a Wilson line becomes irrelevant in the large $N$ limit of M(atrix) theory in which it provides a
light-cone quantization of the original background (without the light-like circle), but it can have large
effects for finite values of $N$. In the case of M theory on $S^1/\Zf_2$, as we discussed above, the theory
compactified on an additional small circle is generally strongly coupled (see \cite{Keurentjes:2006cw} for a
recent discussion), except when we have a Wilson line breaking the $E_8\times E_8$ symmetry to
$\gSO{16}\times\gSO{16}$. The moduli space of this subspace of compactifications on a cylinder is drawn in
figure \ref{modulifigwl}; as can be seen in this figure, the limit of a small space-like circle leads in this
case to a weakly coupled type I string theory (with a Wilson line breaking $\gSO{32}$ to
$\gSO{16}\times\gSO{16}$). The M(atrix) theory for M theory on an interval with this specific light-like Wilson
line is then again a simple theory \cite{Banks:1997zs,Banks:1997it,Lowe:1997sx,Rey:1997hj,Horava:1997ns} -- the
decoupled theory of $N$ D1-branes in this type I background, which is simply an $\grp{SO}{N}$ $1+1$ dimensional
$\Nc=(0,8)$ supersymmetric gauge theory on a circle, coupled to $32$ real left-moving fermions in the
fundamental representation (coming from the D1-D9 strings), half of which are periodic and half of which are
anti-periodic \cite{Polchinski:1995df}.

\FIGURE[h]{
\includegraphics[height=3in]{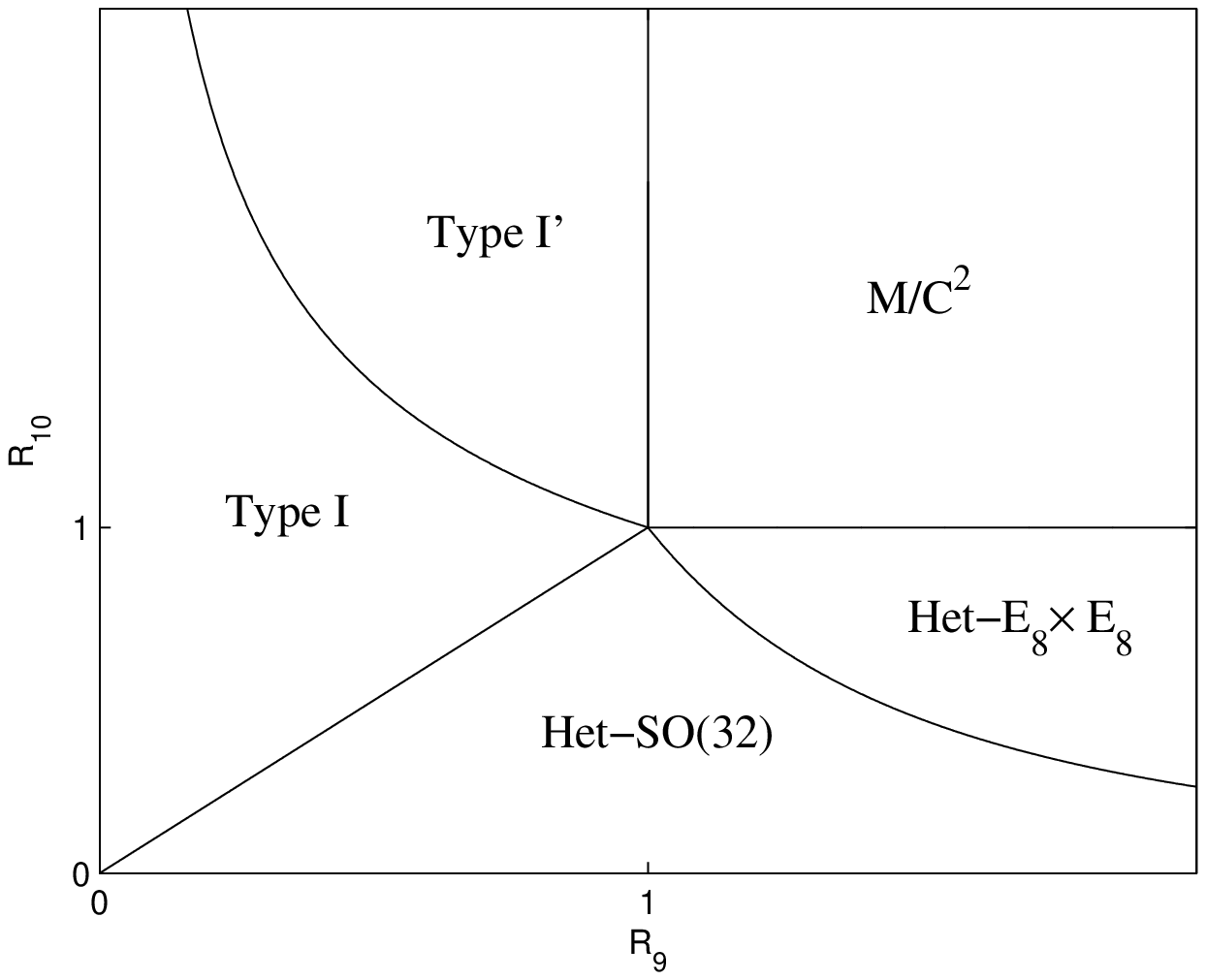}
\caption{The subspace of the moduli space of M theory on a
cylinder, where all backgrounds include a Wilson line breaking the
gauge group to $\gSO{16}\times\gSO{16}$. The parameters in the
plot are the period $R_9$ and length $R_{10}$ of the cylinder in
Planck units.}\label{modulifigwl}}

Now we can move on to the case we are interested in, which is the
M(atrix) theory of M theory on $\Rf^9\times S^1\times(S^1/\Zf_2)$,
with an arbitrary Wilson line $W$ for the $E_8\times E_8$ gauge
group on the $S^1$. To construct the M(atrix) theory we should again
consider the limit of this theory on a very small space-like circle
\cite{Seiberg:1997ad,Sen:1997we}, with a particular scaling of the
size of the cylinder as the size of this extra circle goes to zero.
Again, in general this limit gives a strongly coupled theory, except
in the case where we have an additional Wilson line breaking the
$E_8\times E_8$ gauge group to $\gSO{16}\times\gSO{16}$ on the
additional circle (the original Wilson line $W$ must commute with
this Wilson line in order to obtain a weakly coupled
description).\footnote{In principle we could also have a light-like
Wilson line for the other two $\gU1$ gauge fields appearing in the
low-energy nine dimensional effective action, but we will not
discuss this here.} In such a case we obtain precisely the theory
described in the previous paragraph, compactified on an additional
very small circle with a Wilson line $\tilde{W}$ (which is the
translation of the original Wilson line from the $E_8\times E_8$
variables of the original M theory to the $\gSO{32}$ variables of
the dual type I background). Since the additional circle is very
small we need to perform a T-duality on this circle. We then obtain
a type $I'$ theory of the type described in the previous section,
still compactified on a circle with the $\gSO{16}\times\gSO{16}$
Wilson line, and with the positions of the D8-branes determined by
the eigenvalues of the Wilson line $\tilde{W}$. The D1-branes we had
before now become D2-branes which are stretched both along the
interval between the orientifold planes and along the additional
circle.

The usual derivation of M(atrix) theory
\cite{Seiberg:1997ad,Sen:1997we} shows that the M(atrix) theory is
precisely the decoupled theory living on these D2-branes, in the
limit that the string mass scale goes to infinity keeping the
Yang-Mills coupling constant on the D2-branes, which is
proportional to
\begin{equation}\label{twodg}
g_{YM}^2 \propto g_s / l_s,
\end{equation}
fixed. The D2-brane lives on a cylinder, with a circle of radius $R_1$ (related to the parameters of the
original M theory background by $R_1 = l_p^3 / R_{10} R$ , where $1/R$ is the energy scale associated with the
light-like circle) and an interval of length $\pi R_2$ (given by $R_2 = l_p^3 / R_9 R$). In the standard case of
toroidal compactifications, only the disk contributions to the D2-brane action survive in this limit, giving a
standard supersymmetric Yang-Mills theory, with Yang-Mills coupling $g_{YM}^2 = R / R_9 R_{10}$. However, in our
case it turns out that some contributions to the D2-brane action from M\"obius strip diagrams also survive; this
is evident from the fact that $g_s$ in the type $I'$ background is generally not a constant, leading through
\eqref{twodg} to a non-constant Yang-Mills coupling. This was taken into account in \cite{Kabat:1997za}, where
the Lagrangian for any distribution of D8-branes was obtained, and it was shown that the M\"obius strip
contributions are crucial to cancel anomalies in the gauge theory.

There is one special case when the M\"obius contributions are
absent; this is the case when the type $I'$ background has a
constant dilaton, with eight D8-branes on each orientifold plane.
According to the discussion above, this case provides the DLCQ
description of M theory on a cylinder with a Wilson line breaking
the gauge symmetry to $\gSO{16}\times\gSO{16}$ (so that we are at
some point on the moduli space of figure \ref{modulifigwl}), and
with an additional light-like Wilson line which breaks the gauge
symmetry in the same way. We begin by describing this special case.
In this case the theory on the D2-branes away from the orientifold
planes is just the standard maximally supersymmetric $2+1$
dimensional $\gU{N}$ Yang-Mills theory, with a gauge coupling
related to the parameters of the original M theory background by
$g_{YM}^2 = R / R_9 R_{10}$ (which is the same relation as in
toroidal compactifications). The field content of this theory
includes a gauge field $A_{\mu}$, seven scalar fields $X^j$ and
eight Majorana fermions $\psi_A$. The boundary conditions project
the $\gU{N}$ gauge group to $\grp{SO}{N}$. In addition, the D2-D8
strings give rise to 8 complex chiral fermions in the fundamental
representation $\chi_k$ ($k=1,\cdots,8$) at the boundary $x^2=0$ and
8 additional fermions ${\tilde \chi}_k$ ($k=1,\cdots,8$) at the
other boundary $x^2=\pi R_2$. The action is given by
\begin{multline}\label{cylinder lagrangian s}
    S=\int dt \int_{0}^{2\pi R_1} dx^1\bigg[\int_{0}^{\pi R_2} dx^2
    \frac1{2\gym^2}\tr\bigg(-\half F_{\mu\nu}F^{\mu\nu}+(D_\mu X^j)^2+\\+\half
    [X^j,X^i][X^j,X^i]+i\bar{\psi}_A\gamma^\alpha
    D_\alpha\psi_A-i\bar{\psi}_A\gamma^i_{AB}[X_i,\psi_B]\bigg)+\cr
    +i\sum_{k=1}^{8}
    \bar{\chi}_k(\partial_-+iA_-|_{x^2=0})\chi_k+i\sum_{k=1}^8
    \bar{\tilde{\chi}}_k(\partial_-+iA_-|_{x^2=\pi R_2}){\tilde
    \chi}_k\bigg],
\end{multline}
where $D_{\mu}$ is the covariant derivative for the adjoint
representation, $\partial_-\equiv \partial_t-\partial_1$ and
similarly for $A_-$. Our conventions for fermions and spinor algebra
are summarized in appendix \ref{susyconventions}. Due to the
light-like Wilson line described above, the fermions $\chi_k$ are
periodic around the $x^1$ circle, while the fermions ${\tilde
\chi}_k$ are anti-periodic; this can alternatively be described by
adding a term $\frac{1}{4\pi R_1}$ to the kinetic term of the
${\tilde \chi}_k$ in \eqref{cylinder lagrangian s}.

The boundary conditions can be determined by consistency
conditions for D2-branes ending on an $O8^-$ plane. For the
bosonic fields, at both boundaries, the boundary conditions take
the form
\begin{align}
\nonumber   &X^j=(X^j)^T, &
            &\d_2X^j=-(\d_2X^j)^T, \\
\nonumber   &A^{0,1}=-(A^{0,1})^T, &
            &\d_2A^{0,1}=(\d_2A^{0,1})^T,\\
\label{so conditions s} &A^2= (A^2)^T,\hspace{2em} &
            &\d_2A^2=- (\d_2A^2)^T.
\end{align}
These boundary conditions break the $\gU{N}$ bulk gauge group to
$\gSO{N}$. The zero modes along the interval are $\gSO{N}$ gauge
fields $A^{0,1}$ and eight scalars in the symmetric representation
of $\gSO{N}$ coming from $A^2$ and $X^j$. For the fermions, the
boundary conditions take the form
\begin{align}\label{bc fermions s}
    &\psi_A=-i\gamma^2\psi^T_A,
    &&\d_2\psi_A=i\gamma^2\d_2\psi^T_A.
\end{align}
The zero modes for the right-moving fermions are in the adjoint
representation of $\gSO{N}$, and those of the left-moving fermions
are in the symmetric representation. This leads to an anomaly in
the low-energy $1+1$ dimensional $\gSO{N}$ gauge theory, which is
precisely cancelled by the $16$ additional chiral fermions in the
fundamental representation; this cancellation occurs locally at
each boundary.

The bulk theory has eight $2+1$ dimensional supersymmetries ($16$
real supercharges), but the boundary conditions and the existence
of the D2-D8 fermions break this to a $\Nc=(0,8)$ chiral
supersymmetry (SUSY) in $1+1$ dimensions. The supersymmetry
transformation rules are given by
\begin{align}
\nonumber
    &\delta_\epsilon A_{\alpha}=\frac i2\bar{\epsilon}_A\gamma_\alpha\psi_A,\\
\nonumber
    &\delta_\epsilon
    X^i=-\frac12\bar{\epsilon}_A\gamma^i_{AB}\psi_B,\\
\label{SUSY tranformations}
    &\delta_\epsilon\psi_A=-\frac14F_{\alpha\beta}\gamma^{\alpha\beta}\epsilon_A-\frac
    i2D_{\alpha}X_i\gamma^\alpha\gamma^i_{AB}\epsilon_B-\frac
    i4[X_i,X_j]\gamma^{ij}_{AB}\epsilon_B,\\\nonumber
    &\delta_\epsilon \chi_k=0\,,
    \qquad\delta_\epsilon\tilde\chi_k=0\,.
\end{align}
These transformation rules are consistent with the boundary
conditions \eqref{so conditions s}, \eqref{bc fermions s} only for
\begin{align}\label{epscons}
    \epsilon_A=i\gamma^2\epsilon_A\,,
\end{align}
and thus indeed the boundary conditions preserve only $8$ of the
original $16$ supercharges. Decomposing the fields by their
$\gamma^2$ eigenvalues $\pm i$ :
\begin{align}
    &\epsilon_A=\begin{pmatrix}\epsilon^+_A \\
    \epsilon^-_A\end{pmatrix},
    &&\psi_A=\begin{pmatrix}\psi^+_A \\ \psi^-_A \end{pmatrix},
\end{align}
it follows that the unbroken SUSY is for $\epsilon^-_A$.

In the more general case, as described above, we consider a similar
background but with the D8-branes at arbitrary positions in the
bulk. One obvious change is then that the chiral fermions $\chi$ and
$\tilde\chi$ are no longer localized at the boundaries but rather at
the positions of the D8-branes. More significant changes are that
the varying dilaton leads to a varying gauge coupling constant, and
the background 10-form field in the type $I'$ background leads to a
Chern-Simons term, which is piece-wise constant along the interval.
The most general Lagrangian was written in \cite{Kabat:1997za},
where it was also verified that it is supersymmetric and
anomaly-free. The relation between the positions of the D8-branes
and the varying coupling and 10-form field, which in the bulk
string theory comes from the equations of motion, is reproduced in
the gauge theory by requiring that there is no anomaly in
arbitrary $2+1$ dimensional gauge transformations.

For the purposes of comparison with the Klein bottle case that we will discuss in the next subsection, it is
useful to consider the special case where there is no Wilson loop on the cylinder. This gives, in particular,
the M(atrix) theory of the $E_8\times E_8$ heterotic string compactified on a circle with no Wilson line. In
this case we have a configuration where all D8-branes are on the same orientifold plane, say the one at $x^2=\pi
R_2$. The action in this special case may be written in the form (now denoting the scalar fields by $Y^i$ and
the adjoint fermions by $\Psi_A$)
\begin{multline}\label{right action noa}
    S=\int dt\int_{0}^{2\pi R_1} dx^1
    \bigg[\int_{0}^{\pi R_2} dx^2\frac1{4\gym^2}\tr\biggl(-
    z(x^2)F_{\mu\nu}F^{\mu\nu}+2z^{1/3}(x^2)(D_\mu Y^j)^2+ \\
    +z^{-1/3}(x^2)[Y^j,Y^i][Y^j,Y^i]+2iz^{1/3}(x^2)\bar{\Psi}_A\gamma^\alpha
    D_\alpha\Psi_A+\frac{dz^{1/3}(x^2)}{dx^2}\bar{\Psi}_A\Psi_A-\\-2i\bar{\Psi}_A\gamma^i_{AB}[Y_i,\Psi_B]
    +\frac{4}{3}\frac{dz(x^2)}{dx^2}\e^{\alpha \beta \gamma}(A_\alpha\d_\beta
    A_\gamma+i\frac23 A_\alpha A_\beta A_\gamma)\biggr)+\\
    +i\sum_{k=1}^{8}
    \bar{\chi}_k(\partial_-+iA_-|_{x^2=\pi R_2})\chi_k
   +i\sum_{k=1}^{8}
    \bar{\tilde \chi}_k(\partial_-+iA_-|_{x^2=\pi R_2}){\tilde
    \chi}_k
        \bigg].
\end{multline}
The varying coupling constant is given by
\begin{equation}\label{gymchoice}
    z(x^2)=1+\frac{6\gym^2}{\pi}(x^2 - \frac{\pi R_2}{2}),
\end{equation}
and the coupling grows as we approach the $O8^-$ plane with no
D8-branes on it. Here we arbitrarily defined $\gym$ to be the
effective coupling constant at the middle of the interval (other
choices would modify the constant term in \eqref{gymchoice}). The
linear term in \eqref{gymchoice} is related to the background
10-form field. The effective Yang-Mills coupling constant is
\begin{equation}
(\gym^{eff})^2=\frac{\gym^2}{1+6\gym^2(x^2-\pi
R_2/2)/\pi}=\frac{1}{1/\gym^2+6(x^2/\pi-R_2/2)}.
\end{equation}
This description is valid as long as the Yang-Mills coupling constant does not diverge anywhere, namely for
$g_{YM}^2 < 1 / (3 R_2)$. This is the same condition as the string coupling not diverging in the type $I'$
string theory which we used for deriving this action. The boundary conditions on the fields are essentially the
same as before, but there as some modifications in the boundary conditions which involve derivatives and in the
SUSY transformation laws. The same modifications will appear in the Klein bottle case that we will discuss in
the next subsection, and we will discuss them explicitly there. The fermions $\chi_k$ are still periodic and the
fermions $\tilde{\chi}_k$ anti-periodic due to the light-like Wilson line.

Note that in the two special cases that we described,
\eqref{cylinder lagrangian s} and \eqref{right action noa}, the
theory is exactly free for $N=1$ (as was the original BFSS
M(atrix) theory), since the gauge fields $A_{0,1}$ vanish at both
boundaries; however, this is not the case at more general points
on the moduli space. The usual argument that $N=1$ DLCQ theories
should be free is that $N=1$ is the minimal amount of possible
light-like momentum, so the theory must contain a single particle
with this momentum and no interactions. However, this is no longer
true in the presence of generic light-like Wilson lines, which
modify the quantization of the light-like momentum for charged
states.
%%%%%%%%%%%
%
%%%%%%%%%%%
\subsection{The M(atrix) theory of the Klein bottle compactification}\label{Matrixtheory}

In this subsection we describe the M(atrix) theory of M theory on a
Klein bottle, and we will see that it is very similar to the case
described in the previous subsection.\footnote{The M(atrix) theory
of M theory on a Klein bottle was also discussed in
\cite{Kim:1997aj,Kim:1997uv,Zwart:1997kr} but our results are
different. Perhaps some of these other theories arise from different
choices of light-like Wilson lines.} Again, to derive the M(atrix)
theory we need to consider M theory on a Klein bottle times a very
small space-like circle. Now we do not need to add any Wilson lines
to get a weakly coupled theory; instead we directly obtain $N$
D0-branes in the weakly coupled type IIA string theory on a Klein
bottle that was mentioned in the previous section, in the limit in
which the Klein bottle has a very small size. We then need to
perform two T-dualities to go back to a finite-size compact
manifold. The relevant T-dualities were already described in section
\ref{monk2subsec}: one T-duality (which is straightforward) leads to
the DP background, and the next leads to the $O8^{\pm}$ background.
Thus, the M(atrix) theory is the decoupled theory on $2N$ D2-branes
stretched between an $O8^-$ plane and an $O8^+$ plane\footnote{The
spectrum of D-branes in the $O8^{\pm}$ background was analyzed in
\cite{Bergman:1999ta}.} (we obtain $2N$ D2-branes due to the
presence of D0-branes as well as their images on the original Klein
bottle). This theory is very similar to the theory \eqref{right
action noa} we wrote down in the previous subsection for D2-branes
stretched between an $O8^-$ plane with no D8-branes and another
$O8^-$ plane with 16 D8-branes, since the dilaton and 10-form field
are identical in both of these configurations; the only difference
is that the D2-D8 fermions are not present, and the boundary
conditions on the $O8^+$ plane are different from those on the
$O8^-$ plane. In particular, they project the $\gU{2N}$ gauge group
to $\grp{USp}{2N}$ instead of to $\gSO{2N}$ (which is another way to
see that the rank of the gauge group must be even).

Another naive way to derive this M(atrix) theory would be to start
from the effective action of $N$ D0-branes on the Klein bottle,
and to perform two Fourier transforms of this action, along the
lines of the original derivations of M(atrix) theory
compactifications \cite{Banks:1996vh,Taylor:1996ik}. This analysis
is performed in appendix \ref{QM derivation}; it gives the correct
boundary conditions, but it only gives the terms in the action
coming from the disk and it does not include the effects related
to the variation of $z(x^2)$ which come from M\"obius strip
diagrams, so it leads to an anomalous gauge theory.

The complete action for the M(atrix) theory of M theory on a Klein
bottle is
\begin{multline}\label{right action klein}
    S=\frac1{4\gym^2}\int dt\int_{0}^{2\pi R_1} dx^1\int_{0}^{\pi R_2} dx^2\tr\biggl(-
    z(x^2)F_{\mu\nu}F^{\mu\nu}+2z^{1/3}(x^2)(D_\mu Y^j)^2+ \\
    +z^{-1/3}(x^2)[Y^j,Y^i][Y^j,Y^i]+2iz^{1/3}(x^2)\bar{\Psi}_A\gamma^\alpha
    D_\alpha\Psi_A+\frac{dz^{1/3}(x^2)}{dx^2}\bar{\Psi}_A\Psi_A-\\
    -2i\bar{\Psi}_A\gamma^i_{AB}[Y_i,\Psi_B]
    +\frac{4}{3}\frac{dz(x^2)}{dx^2}\e^{\alpha \beta \gamma}(A_\alpha\d_\beta
    A_\gamma+i\frac23 A_\alpha A_\beta A_\gamma)\biggr)
\end{multline}
where, as in the previous subsection,
\begin{equation}
    z(x^2)=1+\frac{6\gym^2}{\pi}(x^2 - \frac{\pi R_2}{2}).
\end{equation}
The boundary conditions could be derived from the open string theory of D2-branes ending on orientifold planes,
but they can also be derived directly in the gauge theory by requiring the absence of boundary terms and
consistency with the SUSY transformations which are described below. It will be convenient in this section to
think of the $\gU{2N}$ matrices as made of four $N\times N$ blocks, and to use Pauli matrices that are constant
within these blocks. In this notation the scalars $Y^i$ satisfy the boundary conditions\footnote{Note that our
boundary conditions at $x^2=0$ seem different from those of the previous subsection, but the two are simply
related by multiplying all adjoint fields by $\sigma^1$.}
\begin{align}\label{scalars 1}
    &\underline{x^2=0}:
    &&Y^j=\sigma^1(Y^j)^T\sigma^1,
    &&\d_2Y^j=-\sigma^1(\d_2Y^j)^T\sigma^1,\\
 \label{scalars 2}
    &\underline{x^2=\pi R_2}:
    &&Y^j=\sigma^2(Y^j)^T\sigma^2,
    &&\d_2Y^j=-\sigma^2(\d_2Y^j)^T\sigma^2.
\end{align}
The SUSY transformations of the action \eqref{right action klein} are
\begin{align}\label{SUSY1}
    &\delta_\epsilon
    Y^i=-\half\bar{\epsilon}_A\gamma^i_{AB}\psi_B, \\\label{SUSY2}
    &\delta_\epsilon\psi_A=-\frac
    i4z^{-1/3}[Y_i,Y_j]\gamma^{ij}_{AB}\epsilon_B-\frac14z^{1/3}F_{\alpha\beta}\gamma^{\alpha\beta}\epsilon_A
    -\frac i2D_\alpha Y_i\gamma^\alpha\gamma^i_{AB}\epsilon_B,
    \\ \label{SUSY3}
    &\delta_\epsilon A_\alpha=\frac i2
    z^{-1/3}\bar{\epsilon}_A\gamma_\alpha\psi_A.
\end{align}
Unbroken SUSY transformations are those with $\epsilon_A=i\gamma^2\epsilon_A$. Notice that the SUSY
transformations now include the function $z(x^2)$. We can use these transformations to determine the boundary
conditions for the fermions. The non-derivative boundary conditions are the naive ones related to \eqref{scalars
1},\eqref{scalars 2}, namely
\begin{equation}\label{no derivatives fermions}
    \underline{x^2=0} \  : \
    \psi_A=-i\sigma^1\gamma^2\psi^T_A\sigma^1,
    \hspace{3em} \underline{x^2=\pi R_2} \  : \
    \psi_A=-i\sigma^2\gamma^2\psi^T_A\sigma^2.
\end{equation}
The derivative boundary condition for the upper component of the spinor follows immediately from  (\ref{SUSY1}),
\begin{equation}\label{matter multiplet b.c}
    \underline{x^2=0} \  : \
    \d_2\psi^+_A=-\sigma^1(\d_2\psi^+_A)^T\sigma^1,
    \hspace{3em} \underline{x^2=\pi R_2} \  : \
    \d_2\psi^+_A=-\sigma^2(\d_2\psi^+_A)^T\sigma^2.
\end{equation}
For the lower component, we have to use (\ref{SUSY2}) to obtain
\begin{gather}\label{vector multiplet b.c}
     \underline{x^2=0} \  : \ \  \d_2\psi^-_A+\frac13\frac {z'}z\psi^-_A=\sigma^1(\d_2\psi^-_A+\frac13\frac
     {z'}z\psi^-_A)^T\sigma^1,
    \cr \underline{x^2=\pi R_2} \  : \ \
    \d_2\psi^-_A+\frac13\frac {z'}z\psi^-_A=\sigma^2(\d_2\psi^-_A+\frac13\frac
    {z'}z\psi^-_A)^T\sigma^2.
\end{gather}
The deviations from the naive boundary conditions are proportional
to $z'$ which is related to the varying string coupling.

Finally, we will also need to know the boundary conditions on
$A_2$. Note that (\ref{SUSY3}) implies (using
$\bar\epsilon_A=-i\bar\epsilon_A\gamma^2$)
\begin{equation}
    \delta_\epsilon A_2=-\frac 12z^{-1/3}(y)\bar{\epsilon}\psi
    \propto
    z^{-1/3}(y)\epsilon^-\psi^+\,.
\end{equation}
Thus, the boundary conditions on $(z^{1/3}A_2)$ are the same as
those we wrote above for the scalar fields, with no additional
terms. Finally, by further investigation of (\ref{SUSY3}) one can
see that $(z^{1/3}A_{0,1})$ satisfy boundary conditions of exactly
the same form \eqref{vector multiplet b.c} as $\psi^-$.

%%%%%%%%%%%%%%%%%%%%%
%
%
%
%%%%%%%%%%%%%%%%%%%%%
\subsection{The AOA limit of the M(atrix) theory}

As we described in the previous section, there is a limit of M
theory on a Klein bottle, corresponding to small $R_{10}$, which
gives a weakly coupled string theory -- the theory which we called
the AOA background. In this limit we should be able to see that the
M(atrix) theory we constructed becomes a second quantized theory of
strings in this background.\footnote{A similar limit for the theory
described in section \ref{Matcyl} should lead to a second quantized
theory of $E_8\times E_8$ heterotic strings, but we will not discuss
this in detail here.} Recall that the standard M(atrix) theory for
weakly coupled type IIA strings is given by a maximally
supersymmetric $\gU N$ $1+1$ dimensional gauge theory; at low
energies this flows to a sigma model on $\Rf^{8N}/S_N$, which
describes free type IIA strings (written in Green-Schwarz light-cone
gauge)
\cite{Motl:1997th,Banks:1996my,Dijkgraaf:1997vv,Fischler:1997kp}.
The string interactions arise from a twist operator which is the
leading, dimension $3$, correction to the sigma model action
\cite{Dijkgraaf:1997vv}. Similarly, in our case we expect that in
the limit where we should obtain a weakly coupled string theory, the
low-energy effective action should be a symmetric product of the
sigma model of the AOA strings, again written in a Green-Schwarz
light-cone gauge (in this gauge the sigma model action is identical
to that of AOB strings in a static gauge, just like in the type IIA
case we get the action of type IIB strings in a static gauge).

The mapping of parameters described in the previous subsection
implies that the limit of small $R_{10}$ corresponds to small
$R_2$ compared to the other scales in the gauge theory. Thus, in
this limit we obtain the (strongly coupled limit of the) $1+1$
dimensional theory of the zero modes along the interval. We will
analyze this theory in detail for the case of $N=1$, in which the
bulk gauge group is $\gU 2$; it is easy (by a similar analysis to
that of \cite{Dijkgraaf:1997vv}) to see that for higher values of
$N$ we obtain (at low energies) the $N$'th symmetric product of
the $N=1$ theories (deformed by higher dimensional operators
giving the string interactions).

The zero modes for the scalars $Y^i$ are easily determined by
noting that the boundary conditions (\ref{scalars
1}),(\ref{scalars 2}) are satisfied by the identity matrix
\begin{equation}\label{scalars bc}
    Y^i(t,x^1,x^2)_{ab}=Y^i(t,x^1)\mathbb{I}_{ab},
\end{equation}
where $a,b$ are $\gU 2$ indices which we suppress henceforth. The
matrices proportional to the identity matrix are actually a
completely decoupled sector of the theory (for any value of $N$),
with no interactions. The zero mode analysis for the fermions is a
little more involved. The subtlety here is that one should keep in
mind that the fermions are Majorana when deriving the equations of
motion.\footnote{The relevant part of the Lagrangian in components
is
$\mathcal{L}\supseteq-z^{1/3}(\psi^+\d_2\psi^-+\psi^-\d_2\psi^+)+(z^{1/3})'\psi^-\psi^+.$
} To obtain the zero modes we need to solve the equations
\begin{align}\label{equations of motion}
    &z^{1/3}\d_2\psi^+_A=0,
    &&z^{1/3}\d_2\psi^-_A+(z^{1/3})'\psi_A^-=0,
\end{align}
subject to the boundary conditions described in the previous subsection. For the upper component of the spinor
the solution is
\begin{equation*}
    \psi^+_A(t,x^1,x^2)=\psi^+_A(t,x^1) \mathbb{I}\,,
\end{equation*}
which manifestly satisfies the equations of motion and the boundary
conditions.\footnote{The existence of this mode is guaranteed by the
fact that it is actually the Goldstino for the 8 supercharges broken
by the D2-branes.} For the lower component $\psi^-$, the equation of
motion \eqref{equations of motion} guarantees that the derivative
boundary conditions \eqref{vector multiplet b.c} are satisfied. In
order to satisfy the non-derivative boundary conditions \eqref{no
derivatives fermions}, we simply choose the direction in the gauge
group to be $\sigma^3$. Hence, the solution is
\begin{equation}
    \psi^-_A(t,x^1,x^2)=\psi^-_A(t,x^1) z^{-1/3}(x^2) \sigma^3\,.
\end{equation}
Similar zero modes arise for the $A_0$ and $A_1$ component of the gauge field. These lead to a $\gU1$ gauge
field in the low-energy effective action, but since there are no charged fields, this does not lead to any
physical states. Finally, there is a scalar field coming from the zero mode of $A_2$,
\begin{equation}\label{zeromodeA2}
    A_2(t,x^1,x^2)=A_2(t,x^1)z^{-1/3}(x^2) \mathbb{I}\,.
\end{equation}
This scalar is actually compact due to large gauge transformations, as we describe below.

Of course, all these fields fill out $\Nc=(0,8)$ supersymmetry representations in $1+1$ dimensions; the vector
multiplet contains $(A_0,A_1,\psi^-)$, and the matter multiplet contains $(A_2,Y^i,\psi^+)$. For a detailed
description of this kind of SUSY see \cite{Banks:1997zs}. For general values of $N$ we find both types of
multiplet in the adjoint representation of $\gU N$, the same field content as in the M(atrix) theory of type IIA
strings. The low-energy spectrum turns out to be non-chiral (for any value of $N$), guaranteeing that there are
no anomalies.

In order to identify our theory with the AOA background we need to show that the theory is invariant under the
transformation $(-1)^{F_L}$ together with a half-shift on the scalar field coming from $A_2$. Consider a $\gU2$
gauge transformation of the form
\begin{equation}\label{largetrans}
    g(x)=e^{if(x^2)}\sigma_1\,.
\end{equation}
Note that our theory is not invariant under generic $\gU2$ gauge transformations since these are broken by the
boundary conditions. The transformation \eqref{largetrans} preserves all the boundary conditions. In order for
it to leave the theory in the same topological sector (the simplest way to verify this is to regard the cylinder
as an orbifold of the torus and use the usual classification of sectors on the torus) we require that $f(\pi
R_2)-f(0)=(2n+1)\pi/2$ for some integer $n$. In general the transformation \eqref{largetrans} mixes the zero
mode \eqref{zeromodeA2} with other modes of $A_2$; however, all other modes can be gauged away so this mixing is
not really physical. We can work in a gauge where all the non-zero modes are set to zero, and an appropriate
choice of a large gauge transformation which preserves this gauge is
\begin{equation}
    f(x^2)=\frac {z^{2/3}(x^2)\pi/2}{z^{2/3}(\pi R_2 )-z^{2/3}(0)}
    \hspace{1em}\rightarrow\hspace{1em}\d_{2} f(x^2)=\frac{z'}{3}\frac
    {z^{-1/3}(x^2)\pi}{z^{2/3}(\pi R_2 )-z^{2/3}(0)}\,.
\end{equation}
The action of this transformation on the zero mode
\eqref{zeromodeA2} implies that we should identify
\begin{equation}\label{scalaridentification}
    A_2(t,x^1) \simeq A_2(t,x^1)+\frac{1}{3}\frac {z'\pi}{z^{2/3}(\pi R_2
    )-z^{2/3}(0)} = A_2 + \frac{2 g_{YM}^2}{z^{2/3}(\pi R_2
    )-z^{2/3}(0)}.
\end{equation}
In the low-energy effective action, the large gauge transformation
\eqref{largetrans} acts also on the vector multiplet, implying
that the identification \eqref{scalaridentification} is
accompanied by
\begin{equation}
    \psi^-(t,x^1) \rightarrow -\psi^-(t,x^1),\hspace{2em}
    A_{0,1}(t,x^1)\rightarrow-A_{0,1}(t,x^1).
\end{equation}This establishes that our low-energy $1+1$ dimensional sigma model is gauged by ($(-)^{F_L}\times
\text{shift}$), as expected.

Next, we wish to compute the physical radius of the scalar arising from $A_2$ to verify that it agrees with the
physical radius we expect. Carrying out the dimensional reduction explicitly (setting all other fields except
the zero mode of $A_2$ to zero) we get
\begin{multline}\label{physcial radius}
    S=\frac1{4\gym^2}\int d^2x \int_{0}^{\pi R_2}
    dx^2 {\rm Tr}_f\biggl(-
    z(x^2)F_{\mu\nu}F^{\mu\nu}+...\biggr)=\\=
    \frac1{\gym^2}\int d^2x\int_{0}^{\pi R_2} dx^2 ( z^{1/3}(x^2)\d_\mu A_2\d^\mu
    A_2+...)=\\
    =\frac1{\gym^2}\int d^2x(\d_\mu A_2\d^\mu A_2)
    \int_{0}^{\pi R_2} dx^2 z^{1/3}(x^2)=\\=\frac{\pi}{8\gym^4}\int
    d^2x(\d_\mu A_2\d^\mu A_2) (z^{4/3}(\pi R_2)-z^{4/3}(0)).
\end{multline}
Thus, the physical, dimensionless radius of the scalar $A_2$ is
\begin{multline}\label{physicalradiuscorrect}
    \frac{1}{2\pi} \cdot \frac {2\gym^2}{z^{2/3}(\pi R_2 )-z^{2/3}(0)}\cdot \sqrt{\frac{\pi}{8\gym^4}
    (z^{4/3}(\pi R_2)-z^{4/3}(0))}=\\= \frac {1}{ z^{2/3}(\pi
    R_2 )-z^{2/3}(0)}\sqrt{\frac{1}{8\pi}
    (z^{4/3}(\pi R_2)-z^{4/3}(0))}=\\=
    \sqrt{\frac{1}{8\pi}}\sqrt{ \frac{z^{2/3}(\pi R_2)+z^{2/3}(0)}{ z^{2/3}(\pi R_2 )-z^{2/3}(0)}}\,.
\end{multline}
Recall that, as discussed in the previous section, the AOB sigma model has a self-T-duality at a physical radius
of $(8\pi)^{-1/2}$. We see from (\ref{physicalradiuscorrect}) that this corresponds to $z(0)=0$, which is
exactly the case where the Yang-Mills coupling diverges at one side of the interval (due to a diverging coupling
on the $O8^-$ plane in the $O8^{\pm}$ background). As discussed in the previous section, at this point of
diverging coupling the $O8^{\pm}$ background has an enhanced $\gSU2$ gauge symmetry in space-time, which should
correspond to an enhanced $\gSU2$ global symmetry in our gauge theory; we see that in the low-energy effective
action this enhanced global symmetry is precisely the one associated with the AOB sigma model at the self-dual
radius.

In the M(atrix) theory interpretation of our gauge theory, the line
$z(0)=0$ precisely maps to the line of enhanced $\gSU2$ symmetry of
the compactification of M theory on a Klein bottle.
 Note that our gauge theory only makes sense for $z(0) \geq 0$,
since otherwise we obtain negative kinetic terms for some fields.
Thus, our M(atrix) theory description only makes sense above the
self-dual line in figure \ref{modulifigk2}. Of course, the theories
below the line are identified by a duality with the theories above
the line, so we do have a valid description for the full moduli
space of Klein bottle compactifications. A similar analysis for the
case of a cylinder (with no Wilson lines) again shows that infinite
gauge coupling is obtained precisely on the line of enhanced $\gSU2$
symmetry in space-time shown in figure \ref{modulifig16}.

%%%%%%%%%%%%%%%%%%%%%%%%%%%%%%%%%%%%%%%%
%
%   Conclusions
%
%%%%%%%%%%%%%%%%%%%%%%%%%%%%%%%%%%%%%%%%
\section{Conclusions and open questions}\label{Conclusions}
In this paper we analyzed in detail the moduli space of nine
dimensional compactifications of M theory with $\Nc=1$ supersymmetry
and their M(atrix) theory descriptions. We found several surprises :
the moduli space of theories with rank $2$ turned out to have two
disconnected components, and in order to obtain a consistent
description of theories with cross-caps we had to conjecture a
non-perturbative splitting of $O8^0$ planes into a D8-brane and an
infinitely coupled $O8^{(-1)}$ plane. The M(atrix) theories we found
are $2+1$ dimensional gauge theories on a cylinder, but generically
they are rather complicated theories with a varying gauge coupling.
The only case where we obtained a standard gauge theory is the case
of M theory on a cylinder with a Wilson line breaking the gauge
theory to $\gSO{16}\times\gSO{16}$; the simplicity of the M(atrix)
theory in this case is related to the fact that (unlike all other
backgrounds we discussed) this background does not have a subspace
with enhanced gauge symmetry in space-time. In all other cases, the
manifold with enhanced gauge symmetry in space-time is mapped in
M(atrix) theory to having infinite gauge coupling on one side of the
interval.

There are several interesting directions for further research. We
provided some circumstantial evidence for our description of the $X$
and $1/2 X$ backgrounds, but it would be nice to obtain more
evidence for this. One way to obtain such evidence would be to
compactify these backgrounds on an additional circle; these
backgrounds then have an F theory description, with the $O8^+$ plane
becoming an unresolvable $D_8$ singularity \cite{Witten:1997bs}. One
can then consider the nine dimensional limit of this background, as
discussed in \cite{Govindarajan:1997iw,Cachazo:2000ey}, and hope to
recover our picture with the D8-brane emitted into the bulk. Another
possible way to study the $X$ background is by its M(atrix) theory
dual; this is given by the limit of the M(atrix) theory for the
Klein bottle that we constructed in section \ref{Ms} in which the
circle is much smaller than the interval. It would be nice to
understand the dynamics of the theory in this limit in detail in
order to understand the $X$ background better; it may be necessary
for this to analyze the regime of large $N$ with energies of order
$1/N$, which is most directly related to the space-time physics.

Another possible way to study these backgrounds is by brane probes,
such as D2-branes stretched between the two orientifold planes.
These D2-branes are interesting also for another reason, since (at
least naively) they provide the M(atrix) theory for some of the
additional nine dimensional backgrounds that we did not discuss in
the previous section, with the M(atrix) theory for the DP background
(and the other backgrounds in the same moduli space) related to
D2-branes in the $X$ background, and the M(atrix) theory for M
theory on a M\"obius strip related to D2-branes in the $1/2 X$
background. It would be interesting to understand these theories
better; naively one obtains an anomaly from the D2-D8 fermions, and
it is not clear how this is cancelled.

Another natural question involves the realization of the
non-perturbative duality symmetries in the M(atrix) theory. In
toroidal compactifications of M theory, such dualities are related
\cite{Ganor:1996zk} to non-trivial dualities relating electric and
magnetic fields in the M(atrix) theory gauge theories. We mentioned
above that at the enhanced $\gSU{2}$ symmetry points in space-time
the M(atrix) theory is supposed to have a non-trivial enhanced
$\gSU{2}$ global symmetry (which can also be seen by viewing this
theory as the theory of a D2-brane in a type $I'$ background with an
enhanced symmetry); the realization of this symmetry will be
discussed in detail in \cite{future}. More generally, there is a
$\Zf_2$ duality symmetry relating the two sides of each graph in
figures \ref{modulifig16} and \ref{modulifigk2}, which should map to
a duality between two gauge theories in M(atrix) theory.
Unfortunately, so far we have only been able to find the M(atrix)
description for the large radius region of the moduli space (as
described above), and we could not yet find an independent (dual)
description for the small radius region. It would be interesting to
investigate this further \cite{future}; it requires continuing the
moduli space of backgrounds with $O8^-$ planes beyond the point
where the string coupling diverges at one of the orientifold planes,
but without changing to the dual variables.

Finally, it would be interesting to generalize our results
concerning the classification of backgrounds with $16$ supercharges
to lower dimensions (for a partial classification see
\cite{deBoer:2001px}), and to see if there are any new components or
unexplored corners of the moduli space there (as we found in nine
dimensions).

\acknowledgments We would like to thank V. Shpitalnik for
collaboration on the initial stages of this project, and to thank
Y. Antebi, M. Berkooz, S. Hellerman, N. Itzhaki, J. Maldacena, D.
Reichmann, A. Schwimmer and E. Witten for useful discussions. O.A.
and Z.K. would like to thank the Institute for Advanced Study for
its hospitality during the course of this project. This work was
supported in part by the Israel-U.S. Binational Science
Foundation, by the Israel Science Foundation (grant number
1399/04), by the European network HPRN-CT-2000-00122, by a grant
from the G.I.F., the German-Israeli Foundation for Scientific
Research and Development, and by a grant of DIP (H.52).

%%%%%%%%%%%%%%%%%%%%%%%%%%%%%%%%%%%%%%%%
%
%   Appendix
%
%%%%%%%%%%%%%%%%%%%%%%%%%%%%%%%%%%%%%%%%
\appendix

\section{Spinor conventions}\label{susyconventions}

We summarize our conventions for spinors in $2+1$ dimensions. We
choose the metric to be mostly minus, $\eta^{\alpha \beta} = {\rm
diag}(1,-1,-1)$. The gamma matrices of $\gSO{2,1}$ satisfy
$\{\gamma^\alpha,\gamma^\beta\}=2\eta^{\alpha\beta}$, where
$\alpha,\beta$ are vector indices of $\gSO{2,1}$, and the spinor
indices are denoted by $a,b=1,2$ (but are usually suppressed). A
convenient basis for $\gSO{2,1}$ gamma matrices is
\begin{align}
    &\gamma^0=\begin{pmatrix}0&-i\\i&0\end{pmatrix},
    &&\gamma^1=\begin{pmatrix}0&i\\i&0\end{pmatrix},
    &&\gamma^2=\begin{pmatrix}i&0\\0&-i\end{pmatrix}.
\end{align}
Our fermions lie in spinor representations of the global
$Spin(7)_R$ symmetry. The gamma matrices of $Spin(7)_R$ satisfy
$\{\gamma^i,\gamma^j\}=2\delta^{ij}$, where $i$,$j$ are vector
indices of $Spin(7)_R$. We shall denote spinor indices of
$Spin(7)_R$ by $A,B$. Finally, for a Majorana fermion $\psi$, we
define $\bar{\psi}=\psi^T\gamma^0$.

\section{Periodicities of $p$-form fields in some 9d
compactifications}\label{pformrules}

In this appendix we discuss the periodicities of the $p$-form
fields in some of the backgrounds described in section
\ref{monk2subsec}. We being with the AOA background
\cite{Gutperle:2000bf,Hellerman:2005ja}, in which one gauges the
following symmetry of type IIA string theory :
\begin{equation}
    (-)^{F_L}\times (X^9\rightarrow X^9+2\pi R_9).
\end{equation}
The NS-NS 2-form $B_{\mu\nu}$ is periodic along the circle in the
$9$'th direction:
\begin{equation}\label{BfieldHeller}
    B_{\mu\nu}(x^9+2\pi R_9)=B_{\mu\nu}(x^9)
\end{equation}
and leads to a massless mode. On the other hand, the RR 3-form
field $A^{(3)}$ gets a minus sign from $(-)^{F_L}$, hence it has
to be antiperiodic in the direction $x^9$. In components,
\begin{equation}\label{CfieldHeller}
    A^{(3)}_{\mu\nu\rho}(x^9+2\pi R_9)=-A^{(3)}_{\mu\nu\rho}(x^9).
\end{equation}
As discussed in section \ref{monk2subsec}, the target space of the M theory lift of this background is
$\mathbb{R}^{8,1}\times K2$, with $x^{10}\rightarrow -x^{10}$ under the involution. In eleven dimensions the
involution must take $C^{M(3)}\rightarrow -C^{M(3)}$ in order to be a symmetry of M theory. We can also verify
that this is consistent with the periodicities we wrote above. For $C^{M(3)}_{\mu\nu (10)}$, taking into account
the coordinate transformation we get that
\begin{equation}
    C^{M(3)}_{\mu\nu (10)}(x^9+2\pi R_9,-x^{10})=C^{M(3)}_{\mu\nu
    (10)}(x^9,x^{10}).
\end{equation}
Reducing this on $x^{10}$ we find agreement with
(\ref{BfieldHeller}). The other components are all anti-periodic,
in agreement with (\ref{CfieldHeller}).

Next, we compactify this M theory on an additional circle
$\mathbb{R}^{7,1}\times K2\times S^1 $. We can reduce on this
$S^1$ to obtain a different IIA theory, which is the one described
by (\ref{our IIA}). Now the fact that $C^{M(3)}\rightarrow
-C^{M(3)}$ implies different periodicities in the type IIA theory.
For the NS-NS 2-form
\begin{equation}\label{ourIIA Bfield}
    B^{(2)}\rightarrow -B^{(2)},
\end{equation}
which in components implies for instance
\begin{equation}
    B^{(2)}_{\mu 8}(-x^8,x^9+2\pi R_9)=B^{(2)}_{\mu 8}(x^8,x^9).
\end{equation}
Similarly, for the RR 3-form,
\begin{equation}\label{ourIIA Cfield}
    A^{(3)}\rightarrow -A^{(3)}.
\end{equation}
These results can also be derived directly from the IIA perspective.
We first note that $\Omega\times (X^8\rightarrow -X^8)$ is a
symmetry of IIA because it changes the chirality of fermions and the
RR forms twice.\footnote{By super-conformal invariance of the
worldsheet we get that the worldsheet fermions $\psi$ and
$\widetilde{\psi}$ flip sign under this involution and hence change
the chirality in the R sectors.} Since $\Omega$ flips its sign the
$B$ field has to be odd under the involution, in agreement with
(\ref{ourIIA Bfield}). We can check the consistency on the RR sector
in the Green-Schwarz formalism. In this formalism type IIA
superstrings have space-time fermions $\theta_L(z)$ on the left and
$\theta_R(\bar z)$ on the right. The transformation \eqref{our IIA}
that produces the Klein bottle acts on the space-time fermions as
\begin{align}
    &\theta_L(z)\rightarrow \Gamma^8\theta_R(\bar z), \\
    &\theta_R(\bar z)\rightarrow \Gamma^8\theta_L(z).
\end{align}
The RR vertex operator is
\begin{equation}
    V=\theta_R(\bar z)^\dag
    \Gamma_{\mu_1\dotsb\mu_p}\theta_L(z)+\text{c.c.}.
\end{equation}
Under the transformation the fermionic part becomes
\begin{multline}
    \theta_L(z)^\dag\Gamma^8
    \Gamma_{\mu_1\dotsb\mu_p}
    \Gamma^8\theta_R(\bar z)+\text{c.c.}=
    -\theta_R(\bar z)^T(\Gamma^8)^T
    (\Gamma_{\mu_1\dotsb\mu_p})^T
    (\Gamma^8)^T
    \theta_L(z)^*+\text{c.c.}=\\=
    -\theta_R(\bar z)^\dag\Gamma^8
    (\Gamma_{\mu_1\dotsb\mu_p})^\dag
    \Gamma^8\theta_L(z)+\text{c.c.}=
    (-1)^{[p/2]+1}\theta_R(\bar z)^\dag\Gamma^8
    \Gamma_{\mu_1\dotsb\mu_p}
    \Gamma^8\theta_L(z)+\text{c.c.}.
\end{multline}
Thus, the 1-form transforms as a 1-form (changing sign only for the
component in the 8 direction) and the 3-form transforms as a
pseudo-3-form (changing sign for components not in the 8 direction).
In other words, only the twisted 3-form cohomology survives, and
only the untwisted cohomology of 1-forms survives (see
\cite{Hanany:2000fq} for more details\footnote{Loosely speaking, the
untwisted cohomology is the subset of the covering space cohomology
which contains only the forms which are even. The twisted cohomology
contains the odd forms.}).
%%%%%%%%
%
%%%%%%%%
\section{Modular invariance and T-duality in the AOA partition
function}\label{mod and T in partition}

In this appendix we compute the partition function of the AOA theory (defined in section \ref{backgrounds}) and
verify that it is modular invariant and T-duality invariant. Throughout this appendix we set $l_s=1$. The subtle
point is that the definition of the theory includes special phases in the twisted sectors (which are required
for the consistency of the theory).

The sectors of the theory and their matter GSO projections
(including $(-1)^w$) are \cite{Hellerman:2005ja}:\footnote{Note that
the conventions in \cite{Hellerman:2005ja} for right/left-moving
sectors are different from the usual.}
\begin{description}
    \item[A:] $NS-NS$ sector with $n\in\mathbb{Z}, w\in2\mathbb{Z}$
    and GSO projection $-/-$.
    \item[B:] $NS-R$ sector with $n\in\mathbb{Z}, w\in2\mathbb{Z}$
    and GSO projection $-/+$.
    \item[C:] $R-NS$ sector with $n\in\mathbb{Z}+\frac{1}{2}, w\in2\mathbb{Z}$
    and GSO projection $-/-$.
    \item[D:] $R-R$ sector with $n\in\mathbb{Z}+\frac{1}{2}, w\in2\mathbb{Z}$
    and GSO projection $-/+$.
    \item[E:] $R-NS$ sector with $n\in\mathbb{Z}, w\in2\mathbb{Z}+1$
    and GSO projection $+/-$.
    \item[F:] $R-R$ sector with $n\in\mathbb{Z}, w\in2\mathbb{Z}+1$
    and GSO projection $+/+$.
    \item[G:] $NS-NS$ sector with $n\in\mathbb{Z}+\frac{1}{2}, w\in2\mathbb{Z}+1$
    and GSO projection $+/-$.
    \item[H:] $NS-R$ sector with $n\in\mathbb{Z}+\frac{1}{2}, w\in2\mathbb{Z}+1$
    and GSO projection $+/+$.
\end{description}

Let $Z^{\alpha}_{\beta}$ be a path integral on the torus over
fermion fields $\psi$ with periodicities
\begin{gather}
\psi(w+2\pi)=-e^{i\pi\alpha}\psi(w),\cr
\psi(w+2\pi\tau)=-e^{i\pi\beta}\psi(w).
\end{gather}
The fermionic partition sums are then ($0,v,s,c$ stand for the
four possible combinations of NS and R sectors with GSO
projections)
\begin{gather}
    \chi_0= \frac{1}{2}\bigl[Z^0_0(\tau)^4+Z^0_1(\tau)^4\bigr],\cr
    \chi_v= \frac{1}{2}\bigl[Z^0_0(\tau)^4-Z^0_1(\tau)^4\bigr],\cr
    \chi_s= \frac{1}{2}\bigl[Z^1_0(\tau)^4+Z^1_1(\tau)^4\bigr],\cr
    \chi_c= \frac{1}{2}\bigl[Z^1_0(\tau)^4-Z^1_1(\tau)^4\bigr].
\end{gather}
The partition function on the torus that we obtain by summing over
all the sectors described above is :
\begin{multline}\label{Part func}
    Z(\tau)=iV_{10}\int_F\frac{d^2\tau}{16\pi^2\tau_2^2}Z_X^7|\eta(\tau)|^{-2}(\chi_v-\chi_s)
    \times\\\times
    \Biggl(\bar{\chi}_v\sum_{n\in\Zf,w\in2\Zf}\exp\bigl[-\pi\tau_2(\frac{n^2}{R^2}+w^2R^2)+2\pi
    i\tau_1 nw\bigr]\\
    +\bar{\chi}_0 \sum_{n\in\Zf+1/2,w\in 2\Zf+1}\exp\bigl[-\pi\tau_2(\frac{n^2}{R^2}+
    w^2R^2)+2\pi i\tau_1 nw\bigr]\cr -\bar{\chi}_s \sum_{n\in
    \Zf,w\in2\Zf+1}\exp\bigl[-\pi\tau_2(\frac{n^2}{R^2}+w^2R^2)+2\pi
    i\tau_1 nw\bigr]\cr -\bar{\chi}_c \sum_{n\in
    \Zf+1/2,w\in2\Zf}\exp\bigl[-\pi\tau_2(\frac{n^2}{R^2}+w^2R^2)+2\pi
    i\tau_1 nw\bigr] \Biggr),
\end{multline}
where $Z_X^7$ is the partition sum for the $7$ transverse bosons.
After resummation we get
\begin{multline}\label{Part func 2}
    Z(\tau)=iV_{10}\int_F\frac{d^2\tau}{16\pi^2\tau_2^2}Z_X^8(\chi_v-\chi_s) \sum_{m,w \in
    \Zf}\Biggl( \bar{\chi}_v \exp\left[-\frac{\pi
    R^2|m-2w\tau|^2}{\tau_2}\right]\\ +\bar{\chi}_0
    \exp\left[-\frac{\pi R^2|m-(2w+1)\tau|^2}{\tau_2}+\pi
    im\right]\\
    -\bar{\chi}_s \exp\left[-\frac{\pi
    R^2|m-(2w+1)\tau|^2}{\tau_2}\right]\\ -\bar{\chi}_c
    \exp\left[-\frac{\pi R^2|m-2w\tau|^2}{\tau_2}+\pi im\right]
    \Biggr).
\end{multline}
From the modular transformations of $Z^{\alpha}_{\beta}$ :
\begin{gather}
Z^{\alpha}_{\beta}(\tau+1)=\e^{\pi
i(3\alpha^2-1)/12}Z^{\alpha}_{\beta-\alpha+1}(\tau),\cr
Z^{\alpha}_{\beta}(-1/\tau)=Z^{-\beta}_{\alpha},
\end{gather}
we get the transformations of the $\chi$'s :
\begin{gather}
\chi_v(\tau+1)=\e^{2\pi i/3}\chi_v(\tau),\cr
\chi_0(\tau+1)=\e^{-\pi i/3}\chi_0(\tau),\cr
\chi_{s/c}(\tau+1)=\e^{2\pi i/3}\chi_{s/c}(\tau),\cr
\chi_{v/0}(-1/\tau)=\frac{1}{2}\bigl[Z^0_0(\tau)^4\mp
Z^1_0(\tau)^4\bigr],\cr
\chi_{s/c}(-1/\tau)=\frac{1}{2}\bigl[Z^0_1(\tau)^4\pm
Z^1_1(\tau)^4\bigr].
\end{gather}
Modular invariance of \eqref{Part func 2} under
$\tau\rightarrow\tau+1$ is straightforward : the phase from
$(\chi_v-\chi_s)$ is cancelled by the phase coming from the
left-moving sector (the phase of $\bar{\chi}_0$ gets another
$(-1)$ from the transformation of the sum over momenta and
windings). The modular invariance under $\tau\rightarrow
-\frac{1}{\tau}$ follows from the fact that if we take
$m\rightarrow-w$ and $w\rightarrow m$, the bosonic sums that
multiply $(\bar{Z^0_0})^4$ and $(\bar{Z^1_1})^4$ do not change and
the bosonic sums that multiply $(\bar{Z^0_1})^4$ and
$(\bar{Z^1_0})^4$ are switched.

If we look at the partition function (\ref{Part func}), we see
that it is invariant under T-duality with
$R\rightarrow\frac{1}{2R}$, $n\rightarrow \frac{w}{2}$, and
$w\rightarrow 2n$. Recall that in order to verify T-duality one
has to flip the GSO projection of the right-moving modes of the
Ramond sector, and then the partition function is manifestly
invariant (as usual, to verify this one needs to use $Z_1^1=0$).

\section{Quantum mechanics of D0-branes on the Klein bottle}\label{QM derivation}

In this appendix we take a naive approach to the derivation of M(atrix) theory for M theory on a Klein bottle,
by starting with D0-branes on a Klein bottle, imposing the appropriate identifications on the Chan-Paton
indices, and performing a Fourier transform. Our D0-branes are moving in the background with the identification
(\ref{our IIA}), but we will rename the directions of the Klein bottle $x^1$ and $x^2$, and denote the
periodicities of these variables by $2\pi R'_1$ and $4\pi R'_2$. Our identifications of the Chan-Paton indices
will differ from those in \cite{Kim:1997uv} (which describe a non-commutative version, as pointed out in
\cite{Ho:1997yk}), but they are similar to those of \cite{Zwart:1997kr}. We construct the identifications by
simply moving strings with the transformation
\begin{equation}\label{k2 transf}
    (x^1,x^2)\sim(-x^1,x^2+2\pi R'_2)
\end{equation}
and then reversing their orientation. Figure \ref{example
identification} is an example of how we obtain the various
relations.

\begin{figure}[htb]
\begin{center}
\includegraphics[angle=270]{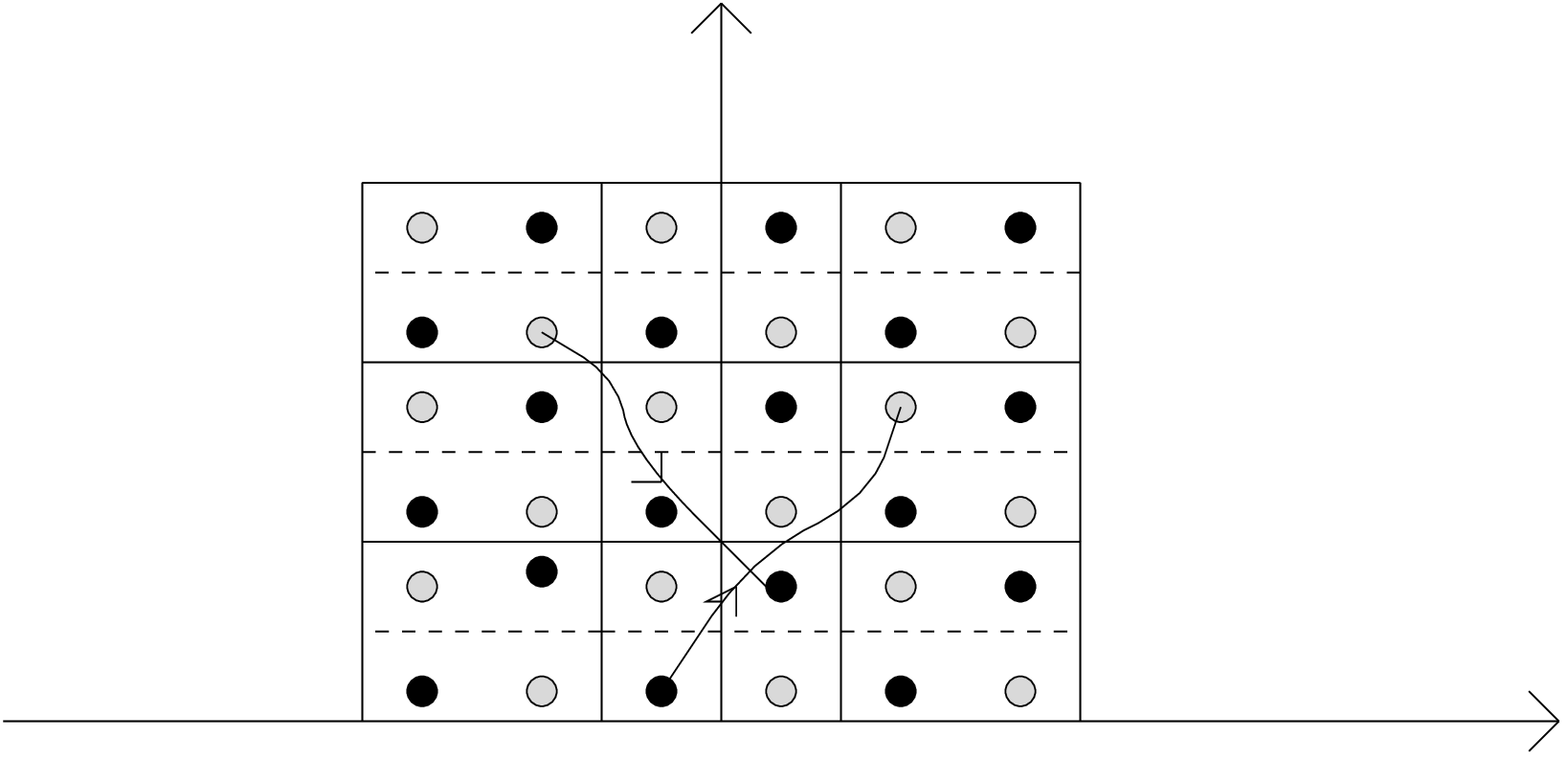}
\caption{To demonstrate the identifications we propose, we show an explicit example of how it works. In this
figure, the middle cell in the lowest row is the (0,0) cell. We act with the orientifold transformation to
demonstrate the following equality $X_{(0,0),(1,1);1,2;i,j}=X_{(-1,2),(0,0);1,2;j,i}=X_{(0,0),(1,-2);1,2;j,i}$
where $X$ is any of the orthogonal coordinates. The dashed line separates each torus fundamental domain to two
copies of the Klein bottle. The filled circles are D0-branes (and the two colors correspond to two different
D0-branes). The two depicted strings are identified under the transformation.}\label{example
identification}\end{center}
\end{figure}

For convenience, we should now have $2N$ D0-branes in each fundamental domain of the torus. Let $X^{i}$  stand
for all the transverse Euclidean directions to the Klein bottle ($i=3,\cdots,9$). We will write everything in
components first and then try to find a more elegant formalism. In general, the indices are
$\Lambda_{k,l;a,b;i,j}$ where $k$ and $l$ are two dimensional vectors of integers (corresponding to the periodic
images of the D0-branes), $a$ and $b$ are in $\{1,2\}$ and $i$ and $j$ define an $N\times N$ matrix. The
identifications coming from the torus periodicities are the standard ones (see, e.g., \cite{Taylor:1996ik}). For
our additional identifications, define $\hat{k}=(-k_1,k_2)$, and then :
\begin{gather}\label{orthogonal}
X^i_{k,l;1,1;i,j}=X^i_{\hat{l},\hat{k};2,2;j,i},\cr
X^i_{k,l;2,2;i,j}=X^i_{\hat{l},\hat{k};1,1;j,i}
=X^i_{\hat{l}+e_2,\hat{k}+e_2;1,1;j,i},\cr
X^i_{k,l;1,2;i,j}=X^i_{\hat{l}+e_2,\hat{k};1,2;j,i},\cr
X^i_{k,l;2,1;i,j}=X^i_{\hat{l},\hat{k}+e_2;2,1;j,i}.\end{gather}
These are completely geometric identifications, similar to the
particular one exhibited in figure \ref{example identification}.
The purpose of writing the second row as it is will become clear
below. One should also notice that there are no identifications on
the orthogonal coordinates besides (\ref{orthogonal}) and the
usual torus identifications.

On the $A^0$ matrices similar identifications hold up to an
additional minus sign because of the action on the vertex operator
(which contains a normal derivative) :
\begin{gather}\label{orthogonal gauge}
A^0_{k,l;1,1;i,j}=-A^0_{\hat{l},\hat{k};2,2;j,i},\cr
A^0_{k,l;2,2;i,j}=-A^0_{\hat{l},\hat{k};1,1;j,i}
=-A^0_{\hat{l}+e_2,\hat{k}+e_2;1,1;j,i},\cr
A^0_{k,l;1,2;i,j}=-A^0_{\hat{l}+e_2,\hat{k};1,2;j,i},\cr
A^0_{k,l;2,1;i,j}=-A^0_{\hat{l},\hat{k}+e_2;2,1;j,i}.
\end{gather}

The constraints on the coordinates $X^1$ and $X^2$ are also very
easy to find by following the fate of a string after the symmetry
operation acts on it. The results are summarized in the following
list :
\begin{gather}\label{parallel}
X^1_{k,l;1,1;i,j}=-X^1_{\hat{l},\hat{k};2,2;j,i},\cr
X^1_{k,l;2,2;i,j}=-X^1_{\hat{l},\hat{k};1,1;j,i}=-X^1_{\hat{l}+e_2,\hat{k}+e_2;1,1;j,i},\cr
X^1_{k,l;1,2;i,j}=-X^1_{\hat{l}+e_2,\hat{k};1,2;j,i}, \cr
X^1_{k,l;2,1;i,j}=-X^1_{\hat{l},\hat{k}+e_2;2,1;j,i},\cr
X^2_{k,l;1,1;i,j}=X^2_{\hat{l},\hat{k};2,2;j,i}-\delta_{l,k}2\pi
R'_2\delta_{i,j}, \cr
X^2_{k,l;2,2;i,j}=X^2_{\hat{l},\hat{k};1,1;j,i}+\delta_{l,k}2\pi
R'_2\delta_{i,j}=X^2_{\hat{l}+e_2,\hat{k}+e_2;1,1;j,i}-\delta_{l,k}2\pi
R'_2\delta_{i,j}, \cr
X^2_{k,l;1,2;i,j}=X^2_{\hat{l}+e_2,\hat{k};1,2;j,i}, \cr
X^2_{k,l;2,1;i,j}=X^2_{\hat{l},\hat{k}+e_2;2,1;j,i}.
\end{gather}

As was explained in section \ref{Matrixtheory}, we know there
exists a nice T-dual description of this theory as the worldvolume
of a D2-brane. It should be possible, therefore, to rearrange the
very inconvenient set of identifications on the infinite matrices
that we wrote above as a $2+1$ dimensional gauge theory of finite
matrices.

To see this, define
\begin{equation}
    M_{k,k';a,a'}=\delta_{k',\hat{k}+\tau_{a,a'}e_2}\sigma^1_{a,a'},
\end{equation}
where we have defined
\begin{equation}
\tau=\half\left(\sigma^1-i\sigma^2\right)=\begin{pmatrix}0 & 0
\\ 1 & 0\end{pmatrix}.
\end{equation}
One can easily show that
\begin{equation}
    (M^{-1})_{k',l;a',a}=\delta_{k',\hat{l}+\tau_{a,a'}e_2}\sigma^1_{a',a}.
\end{equation}
Now, equations (\ref{orthogonal}),(\ref{orthogonal gauge}),(\ref{parallel}) can be neatly rewritten in the
following form :
\begin{gather}\label{matrix rewriting} X^i=M(X^i)^TM^{-1},\cr
A^0=-M(A^0)^TM^{-1},\cr X^1=-M(X^1)^TM^{-1},\cr
X^2=M(X^2)^TM^{-1}-2\pi R'_2\delta_{k,l}\delta_{a,b}\delta_{i,j}.
\end{gather}
Of course, these identifications are imposed together with the
torus identifications. The equivalence of (\ref{matrix rewriting})
with (\ref{orthogonal}),(\ref{orthogonal gauge}),(\ref{parallel})
can be demonstrated by simple calculations.

Next, we rearrange the fields as Fourier components in the $k$,$l$
indices :
\begin{gather}
A^0_{a,b;i,j}(x,t)=\sum_k (A^0)_{0,k;a,b;i,j}e^{i\left(\frac{k_1
x_1}{R_1}+\frac{k_2 x_2}{R_2}\right)}, \cr
A^1_{a,b;i,j}(x,t)=\sum_k X^1_{0,k;a,b;i,j}e^{i\left(\frac{k_1
x_1}{R_1}+\frac{k_2 x_2}{R_2}\right)},\cr
A^2_{a,b;i,j}(x,t)=\sum_k
X^2_{0,k;a,b;i,j}e^{i\left(\frac{k_1x_1}{R_1}+\frac{k_2
x_2}{R_2}\right)},\cr \forall l=3...9  \ \
X^l_{a,b;i,j}(x,t)=\sum_k X^l_{0,k;a,b;i,j}e^{i\left(\frac{k_1
x_1}{R_1}+\frac{k_2 x_2}{R_2}\right)},
\end{gather}
where $R_1\equiv l_s^2/R'_1$ and $R_2\equiv l_s^2/2R'_2$ . The
Lagrangian governing these fields is given by a maximally
supersymmetric (after including the fermions) Yang-Mills theory in
$2+1$ dimensions ($16$ real supercharges), which is the dimensional
reduction of the ten dimensional $\mathcal{N}=1$ SYM theory.
However, there are some inter-relations among the fields which we
now derive. For convenience we suppress the $i,j$ indices. The
inter-relations comprise the following set of additional relations
on the theory with $16$ supercharges ($j=3,\cdots,9$) :
\begin{equation}\label{cylinder identifications}
\begin{split}
    &X^j_{1,1}(x,t)=(X^j_{2,2}(-\hat x,t))^T,\\
    &X^j_{1,2}(x,t)=e^{-i\frac{x_2}{R_2}}(X^j_{1,2}(-\hat x,t))^T, \\
    &X^j_{2,1}(x,t)=e^{i\frac{x_2}{R_2}}(X^j_{2,1}(-\hat x,t))^T,\\
    &A^0_{1,1}(x,t)=-(A^0_{2,2}(-\hat x,t))^T,\\
    &A^0_{1,2}(x,t)=-e^{-i\frac{x_2}{R_2}}(A^0_{1,2}(-\hat
    x,t))^T,
    \\
    &A^0_{2,1}(x,t)=-e^{i\frac{x_2}{R_2}}(A^0_{2,1}(-\hat x,t))^T, \\
    &A^1_{1,1}(x,t)=-(A^1_{2,2}(-\hat x,t))^T,\\
    &A^1_{1,2}(x,t)=-e^{-i\frac{x_2}{R_2}}(A^1_{1,2}(-\hat
    x,t))^T,
    \\
    &A^1_{2,1}(x,t)=-e^{i\frac{x_2}{R_2}}(A^1_{2,1}(-\hat x,t))^T, \\
    &A^2_{1,1}(x,t)=(A^2_{2,2}(-\hat x,t))^T-2\pi R'_2\delta_{i,j},\\
    &A^2_{1,2}(x,t)=e^{-i\frac{x_2}{R_2}}(A^2_{1,2}(-\hat x,t))^T,\\
    &A^2_{2,1}(x,t)=e^{i\frac{x_2}{R_2}}(A^2_{2,1}(-\hat x,t))^T.
\end{split}
\end{equation}

We see that our theory can be written on a cylinder of volume $(2\pi R_1)\times (\pi R_2)$ with orientifold
planes at the boundaries. We can read the relevant boundary conditions from (\ref{cylinder identifications}). As
the action appears to be an orbifold of the maximally supersymmetric action we write it explicitly :
\begin{equation}S=\frac1{2\gym^2}\int_{\mathbb{T}^2} dtd^2x
{\rm Tr}\left(-\half F_{\mu\nu}F^{\mu\nu}+(D_\mu X^j)^2+\half
[X^j,X^i][X^j,X^i]+\text{fermions}\right).
\end{equation}
The size of $\mathbb{T}^2$ is $(2\pi R_1)\times (2\pi R_2)$. All
the fields are periodic and in our case take the form
\begin{equation}
X=\begin{pmatrix}X_{1,1;i,j} & X_{1,2;i,j} \\ X_{2,1;i,j}
&X_{2,2;i,j}
\end{pmatrix}, \hspace{3em} A=\begin{pmatrix}A_{1,1;i,j}
& A_{1,2;i,j} \\ A_{2,1;i,j} &A_{2,2;i,j}
\end{pmatrix}, \hspace{3em} i,j=1,\cdots,N.
\end{equation}
We split the action :
\begin{multline}\label{torus lagrangian '}
    S=\frac1{2\gym^2}\int dt \int_{0}^{2\pi R_1} dx^1\int_{-\pi R_2}^{0}
    dx^2\tr\left(-\half F_{\mu\nu}F^{\mu\nu}+(D_\mu X^j)^2+\half
    [X^j,X^i][X^j,X^i]\right)+\\
    +\frac1{2\gym^2}\int dt \int_{0}^{2\pi
    R_1} dx^1\int_{0}^{\pi R_2} dx^2\tr\left(-\half
    F_{\mu\nu}F^{\mu\nu}+(D_\mu X^j)^2+\half
    [X^j,X^i][X^j,X^i]\right)
\end{multline}

Using (\ref{cylinder identifications}) we can reexpress all the
fields in the domain $[-\pi R_2,0]$ as fields in the domain
$[0,\pi R_2]$. The computation becomes very easy if one notices
that the set of equations (\ref{cylinder identifications}) is
actually simple, $\forall j=3,\cdots,9$ :
\begin{gather}\label{simple cylinder ident1}
    X^j(x,t)=e^{-\frac{ix^2}{2R_2}\sigma^3}\sigma^1(X^j(-\hat{x},t))^T\sigma^1e^{\frac{ix^2}{2R_2}\sigma^3},\cr
    A^{0(1)}(x,t)=-e^{-\frac{ix^2}{2R_2}\sigma^3}\sigma^1(A^{0(1)}(-\hat{x},t))^T\sigma^1e^{\frac{ix^2}{2R_2}\sigma^3},\cr
    A^{2}(x,t)=e^{-\frac{ix^2}{2R_2}\sigma^3}\sigma^1(A^{2}(-\hat{x},t))^T\sigma^1e^{\frac{ix^2}{2R_2}\sigma^3}-2\pi
    R'_2\sigma^3,
\end{gather}
where the relations are on the full $2N\times 2N$ matrices. The
sigma matrices are always tensored with the unit matrix of
dimension $N\times N$. One can see that this gives a nice
Lagrangian, as most of the commutators are invariant, or change
sign. One easy way to proceed is to rewrite the above lagrangian
with new fields in the following way :
\begin{gather}\label{gauge transformation} \wt
X^j(x,t)=e^{\frac{ix^2\sigma^3}{4R_2}}X^j(x,t)e^{-\frac{ix^2\sigma^3}{4R_2}},
\cr \wt
A^{0(1)}(x,t)=e^{\frac{ix^2\sigma^3}{4R_2}}A^{0(1)}(x,t)e^{-\frac{ix^2\sigma^3}{4R_2}},
\cr \wt A^2(x,t)=
e^{\frac{ix^2\sigma^3}{4R_2}}A^{2}(x,t)e^{-\frac{ix^2\sigma^3}{4R_2}}+\sigma^3
\pi R'_2.
\end{gather}
This transformation is a $\gU{2N}$ gauge transformation. Hence the
Lagrangian of the tilded fields is actually the same Lagrangian as
(\ref{torus lagrangian '}). We will still name those fields in the
usual notation omitting the tildes. The identification of the
tilded fields\footnote{As mentioned, we won't use the notation of
the tildes anymore.} is very suggestive:
\begin{gather}\label{simple cylinder ident2}
    X^j(x,t)=\sigma^1(X^j(-\hat{x},t))^T\sigma^1,\cr
    A^{0(1)}(x,t)=-\sigma^1(A^{0(1)}(-\hat{x},t))^T\sigma^1,\cr
    A^2(x,t)=\sigma^1 (A^2(-\hat x,t))^T\sigma^1.
\end{gather}
Using this transformations we transform the $[-\pi R_2,0]$ interval of the action to $[0,\pi R_2]$, and we find
that it precisely agrees with the original action for this interval. The relative minus signs in the gauge
fields exactly compensate for the minus signs that arise because $[A^T,B^T]=-([A,B])^T$ and because
$\partial_2\rightarrow-\partial_2$. So, our final result is
\begin{multline}\label{cylinder lagrnangian}S=\frac1{2\gym^2}\int dt
\int_{0}^{2\pi R_1} dx^1\int_{0}^{\pi R_2} dx^2\tr\Big(-\half
F_{\mu\nu}F^{\mu\nu}+(D_\mu X^j)^2+\\+\half
[X^j,X^i][X^j,X^i]+\text{fermions}\Big).\end{multline}
We will determine $\gym$ soon, so we omitted the factor 2 for
now.\footnote{The Lagrangian (\ref{cylinder lagrnangian}) on the
cylinder defines what we mean by $\gym$ so there is no ambiguity.}
The boundary conditions are interesting. Equations (\ref{gauge
transformation}) tell us that
\begin{gather}
    X^j(x^1,x^2+2\pi R_2,t)=\sigma^3X^j(x^1,x^2,t)\sigma^3,
    \cr A^{0,1,2}(x^1,x^2+2\pi
    R_2,t)=\sigma^3A^{0,1,2}(x^1,x^2,t)\sigma^3.
\end{gather}
Combined with equations (\ref{simple cylinder ident2}) we obtain
\begin{gather}\label{twisted boundary}
    X^j(x^1,-x^2+2\pi R_2,t)=\sigma^2(X^j(x,t))^T\sigma^2,
    \cr A^{0,1}(x^1, -x^2+2\pi R_2,t)=-\sigma^2(A^{0,1}(x^1,x^2,t))^T\sigma^2,\cr
    A^{2}(x^1,-x^2+2\pi R_2,t)=\sigma^2(A^{2}(x^1,
    x^2,t))^T\sigma^2.
\end{gather}
The boundary conditions at the two orientifold planes are, hence,
different.\footnote{These are the anticipated $O^-$ and $O^+$
planes.}
 For $x^2=0$  the matrices satisfy
\begin{align}\label{boundary conditions}
    &X^j=\sigma^1(X^j)^T\sigma^1,
    &&\d_2X^j=-\sigma^1(\d_2X^j)^T\sigma^1,\\
    &A^{0(1)}=-\sigma^1(A^{0(1)})^T\sigma^1,
    &&\d_2A^{0(1)}=\sigma^1(\d_2A^{0(1)})^T\sigma^1,\\
    &A^2=\sigma^1 (A^2)^T\sigma^1,
    &&\d_2A^2=-\sigma^1 (\d_2A^2)^T\sigma^1,
\end{align}
so they are of the form
\begin{gather}A^{2} \ , \ X^j=\begin{pmatrix}N & S \\ \tilde S & N^T\end{pmatrix}\hspace{2em} N^+=N ,\ S^T=S  ,\ \tilde
S^T=\tilde S ,\ S^+=\tilde S,
 \cr
  A^{0} \ , \ A^{1} = \begin{pmatrix}M & R \\ \tilde R & -M^T\end{pmatrix}\hspace{2em} M^+=M ,\ R^T=-R  ,\ \tilde
R^T=-\tilde R ,\ R^+=\tilde R.   \end{gather}

Collecting all these facts together we get that the gauge group in
the bulk is $\gU{2N}$ while on the $x^2=0$ boundary the gauge
group is $\gSO{2N}$. On this boundary the fields $A^2$, $X^j$ are
in the symmetric representation of the gauge group.

The second boundary $x^2=\pi R_2$ is substantially different. We
write the boundary conditions there in a familiar form
\begin{align}\label{twisted boundary conditions}
    &\sigma^2X^j\sigma^2=(X^j)^T,
    &&\sigma^2 \d_2X^j\sigma^2=-(\d_2X^j)^T,\\
    &\sigma^2A^{0,1}\sigma^2=-(A^{0,1})^T,
    &&\sigma^2\d_2A^{0,1}\sigma^2=(\d_2A^{0,1})^T,\\
    &\sigma^2A^{2}(x^1,t)\sigma^2=(A^{2}(x^1,t))^T,
    &&\sigma^2\d_2A^{2}\sigma^2=-(\d_2A^{2})^T.
\end{align}
Consequently, the gauge group at this boundary is $\grp{USp}{2N}$.
The most general matrices are of the form
\begin{gather}A^{2} \ , \ X^j=\begin{pmatrix}N & S \\ \tilde S & N^T\end{pmatrix}\hspace{2em} N^+=N ,\ S^T=-S  ,\ \tilde
S^T=-\tilde S ,\ S^+=\tilde S,
 \cr
  A^{0} \ , \ A^{1} = \begin{pmatrix}M & R \\ \tilde R & -M^T\end{pmatrix}\hspace{2em} M^+=M ,\ R^T=R  ,\ \tilde
R^T=\tilde R ,\ R^+=\tilde R.   \end{gather}

It is easy to verify that this precisely agrees with the action and
boundary conditions that we wrote down in the main text for the
D2-brane between two orientifold planes, if we only include the disk
contributions to this action and not the M\"obius strip
contributions.

%%%%%%%%%%%%%%%%%%%%%%%%%%%%%%%%%%%%%%%%%%%%%%%%%%%%%%%%%%%%%%%%%%%%%%%%%%%%%%%%%%%%%%%%%%%%%%%%
%Included for Gather Purpose only:
%input "klein.bib"
\bibliographystyle{jhep}
\bibliography{klein}

\providecommand{\href}[2]{#2}\begingroup\raggedright\begin{thebibliography}{10}

\bibitem{deBoer:2001px}
J.~de~Boer, R.~Dijkgraaf, K.~Hori, A.~Keurentjes, J.~Morgan, D.~R. Morrison,
  and S.~Sethi, {\it Triples, fluxes, and strings},  {\em Adv. Theor. Math.
  Phys.} {\bf 4} (2002) 995--1186,
  [\href{http://xxx.lanl.gov/abs/hep-th/0103170}{{\tt hep-th/0103170}}].

\bibitem{Keur-private}
A.~Keurentjes.
\newblock Private communication.

\bibitem{Kabat:1997za}
D.~Kabat and S.~J. Rey, {\it Wilson lines and {T}-duality in heterotic
  {M}(atrix) theory},  {\em Nucl. Phys.} {\bf B508} (1997) 535--568,
  [\href{http://xxx.lanl.gov/abs/hep-th/9707099}{{\tt hep-th/9707099}}].

\bibitem{Horava:1995qa}
P.~Horava and E.~Witten, {\it Heterotic and type {I} string dynamics from
  eleven dimensions},  {\em Nucl. Phys.} {\bf B460} (1996) 506--524,
  [\href{http://xxx.lanl.gov/abs/hep-th/9510209}{{\tt hep-th/9510209}}].

\bibitem{Polchinski:1995df}
J.~Polchinski and E.~Witten, {\it Evidence for heterotic - type {I} string
  duality},  {\em Nucl. Phys.} {\bf B460} (1996) 525--540,
  [\href{http://xxx.lanl.gov/abs/hep-th/9510169}{{\tt hep-th/9510169}}].

\bibitem{Dai:1989ua}
J.~Dai, R.~G. Leigh, and J.~Polchinski, {\it New connections between string
  theories},  {\em Mod. Phys. Lett.} {\bf A4} (1989) 2073--2083.

\bibitem{Morrison:1996xf}
D.~R. Morrison and N.~Seiberg, {\it Extremal transitions and five-dimensional
  supersymmetric field theories},  {\em Nucl. Phys.} {\bf B483} (1997)
  229--247, [\href{http://xxx.lanl.gov/abs/hep-th/9609070}{{\tt
  hep-th/9609070}}].

\bibitem{Bergman:1997py}
O.~Bergman, M.~R. Gaberdiel, and G.~Lifschytz, {\it String creation and
  heterotic-type {I}' duality},  {\em Nucl. Phys.} {\bf B524} (1998) 524--544,
  [\href{http://xxx.lanl.gov/abs/hep-th/9711098}{{\tt hep-th/9711098}}].

\bibitem{Bachas:1997kn}
C.~P. Bachas, M.~B. Green, and A.~Schwimmer, {\it (8,0) quantum mechanics and
  symmetry enhancement in type {I}' superstrings},  {\em JHEP} {\bf 01} (1998)
  006, [\href{http://xxx.lanl.gov/abs/hep-th/9712086}{{\tt hep-th/9712086}}].

\bibitem{Dabholkar:1996pc}
A.~Dabholkar and J.~Park, {\it Strings on orientifolds},  {\em Nucl. Phys.}
  {\bf B477} (1996) 701--714,
  [\href{http://xxx.lanl.gov/abs/hep-th/9604178}{{\tt hep-th/9604178}}].

\bibitem{Keurentjes:2000bs}
A.~Keurentjes, {\it Orientifolds and twisted boundary conditions},  {\em Nucl.
  Phys.} {\bf B589} (2000) 440--460,
  [\href{http://xxx.lanl.gov/abs/hep-th/0004073}{{\tt hep-th/0004073}}].

\bibitem{Gutperle:2000bf}
M.~Gutperle, {\it Non-{BPS} {D}-branes and enhanced symmetry in an asymmetric
  orbifold},  {\em JHEP} {\bf 08} (2000) 036,
  [\href{http://xxx.lanl.gov/abs/hep-th/0007126}{{\tt hep-th/0007126}}].

\bibitem{Hellerman:2005ja}
S.~Hellerman, {\it New type {II} string theories with sixteen supercharges},
  \href{http://xxx.lanl.gov/abs/hep-th/0512045}{{\tt hep-th/0512045}}.

\bibitem{Vafa:1995gm}
C.~Vafa and E.~Witten, {\it Dual string pairs with {N}=1 and {N}=2
  supersymmetry in four dimensions},  {\em Nucl. Phys. Proc. Suppl.} {\bf 46}
  (1996) 225--247, [\href{http://xxx.lanl.gov/abs/hep-th/9507050}{{\tt
  hep-th/9507050}}].

\bibitem{Witten:1997bs}
E.~Witten, {\it Toroidal compactification without vector structure},  {\em
  JHEP} {\bf 02} (1998) 006,
  [\href{http://xxx.lanl.gov/abs/hep-th/9712028}{{\tt hep-th/9712028}}].

\bibitem{Keurentjes:2001cp}
A.~Keurentjes, {\it Discrete moduli for type {I} compactifications},  {\em
  Phys. Rev.} {\bf D65} (2002) 026007,
  [\href{http://xxx.lanl.gov/abs/hep-th/0105101}{{\tt hep-th/0105101}}].

\bibitem{Sen:1996vd}
A.~Sen, {\it {F}-theory and orientifolds},  {\em Nucl. Phys.} {\bf B475} (1996)
  562--578, [\href{http://xxx.lanl.gov/abs/hep-th/9605150}{{\tt
  hep-th/9605150}}].

\bibitem{Chaudhuri:1995fk}
S.~Chaudhuri, G.~Hockney, and J.~D. Lykken, {\it Maximally supersymmetric
  string theories in {D}<10},  {\em Phys. Rev. Lett.} {\bf 75} (1995)
  2264--2267, [\href{http://xxx.lanl.gov/abs/hep-th/9505054}{{\tt
  hep-th/9505054}}].

\bibitem{Chaudhuri:1995bf}
S.~Chaudhuri and J.~Polchinski, {\it Moduli space of {CHL} strings},  {\em
  Phys. Rev.} {\bf D52} (1995) 7168--7173,
  [\href{http://xxx.lanl.gov/abs/hep-th/9506048}{{\tt hep-th/9506048}}].

\bibitem{Park:1996it}
J.~Park, {\it Orientifold and {F}-theory duals of {CHL} strings},  {\em Phys.
  Lett.} {\bf B418} (1998) 91--97,
  [\href{http://xxx.lanl.gov/abs/hep-th/9611119}{{\tt hep-th/9611119}}].

\bibitem{Mikhailov:1998si}
A.~Mikhailov, {\it Momentum lattice for {CHL} string},  {\em Nucl. Phys.} {\bf
  B534} (1998) 612--652, [\href{http://xxx.lanl.gov/abs/hep-th/9806030}{{\tt
  hep-th/9806030}}].

\bibitem{Banks:1996vh}
T.~Banks, W.~Fischler, S.~H. Shenker, and L.~Susskind, {\it M theory as a
  matrix model: A conjecture},  {\em Phys. Rev.} {\bf D55} (1997) 5112--5128,
  [\href{http://xxx.lanl.gov/abs/hep-th/9610043}{{\tt hep-th/9610043}}].

\bibitem{Susskind:1997cw}
L.~Susskind, {\it Another conjecture about m(atrix) theory},
  \href{http://xxx.lanl.gov/abs/hep-th/9704080}{{\tt hep-th/9704080}}.

\bibitem{Seiberg:1997ad}
N.~Seiberg, {\it Why is the matrix model correct?},  {\em Phys. Rev. Lett.}
  {\bf 79} (1997) 3577--3580,
  [\href{http://xxx.lanl.gov/abs/hep-th/9710009}{{\tt hep-th/9710009}}].

\bibitem{Sen:1997we}
A.~Sen, {\it D0 branes on {$T^n$} and matrix theory},  {\em Adv. Theor. Math.
  Phys.} {\bf 2} (1998) 51--59,
  [\href{http://xxx.lanl.gov/abs/hep-th/9709220}{{\tt hep-th/9709220}}].

\bibitem{Keurentjes:2006cw}
A.~Keurentjes, {\it Determining the dual},
  \href{http://xxx.lanl.gov/abs/hep-th/0607069}{{\tt hep-th/0607069}}.

\bibitem{Banks:1997zs}
T.~Banks, N.~Seiberg, and E.~Silverstein, {\it Zero and one-dimensional probes
  with {N}=8 supersymmetry},  {\em Phys. Lett.} {\bf B401} (1997) 30--37,
  [\href{http://xxx.lanl.gov/abs/hep-th/9703052}{{\tt hep-th/9703052}}].

\bibitem{Banks:1997it}
T.~Banks and L.~Motl, {\it Heterotic strings from matrices},  {\em JHEP} {\bf
  12} (1997) 004, [\href{http://xxx.lanl.gov/abs/hep-th/9703218}{{\tt
  hep-th/9703218}}].

\bibitem{Lowe:1997sx}
D.~A. Lowe, {\it Heterotic matrix string theory},  {\em Phys. Lett.} {\bf B403}
  (1997) 243--249, [\href{http://xxx.lanl.gov/abs/hep-th/9704041}{{\tt
  hep-th/9704041}}].

\bibitem{Rey:1997hj}
S.-J. Rey, {\it Heterotic {M}(atrix) strings and their interactions},  {\em
  Nucl. Phys.} {\bf B502} (1997) 170--190,
  [\href{http://xxx.lanl.gov/abs/hep-th/9704158}{{\tt hep-th/9704158}}].

\bibitem{Horava:1997ns}
P.~Horava, {\it Matrix theory and heterotic strings on tori},  {\em Nucl.
  Phys.} {\bf B505} (1997) 84--108,
  [\href{http://xxx.lanl.gov/abs/hep-th/9705055}{{\tt hep-th/9705055}}].

\bibitem{Kim:1997aj}
N.~Kim and S.-J. Rey, {\it Non-orientable {M}(atrix) theory},
  \href{http://xxx.lanl.gov/abs/hep-th/9710192}{{\tt hep-th/9710192}}.

\bibitem{Kim:1997uv}
N.~Kim and S.~J. Rey, {\it {M}(atrix) theory on an orbifold and twisted
  membrane},  {\em Nucl. Phys.} {\bf B504} (1997) 189--213,
  [\href{http://xxx.lanl.gov/abs/hep-th/9701139}{{\tt hep-th/9701139}}].

\bibitem{Zwart:1997kr}
G.~Zwart, {\it Matrix theory on non-orientable surfaces},  {\em Phys. Lett.}
  {\bf B429} (1998) 27--34, [\href{http://xxx.lanl.gov/abs/hep-th/9710057}{{\tt
  hep-th/9710057}}].

\bibitem{Bergman:1999ta}
O.~Bergman, E.~G. Gimon, and P.~Horava, {\it Brane transfer operations and
  t-duality of non-bps states},  {\em JHEP} {\bf 04} (1999) 010,
  [\href{http://xxx.lanl.gov/abs/hep-th/9902160}{{\tt hep-th/9902160}}].

\bibitem{Taylor:1996ik}
W.~Taylor, {\it D-brane field theory on compact spaces},  {\em Phys. Lett.}
  {\bf B394} (1997) 283--287,
  [\href{http://xxx.lanl.gov/abs/hep-th/9611042}{{\tt hep-th/9611042}}].

\bibitem{Motl:1997th}
L.~Motl, {\it Proposals on nonperturbative superstring interactions},
  \href{http://xxx.lanl.gov/abs/hep-th/9701025}{{\tt hep-th/9701025}}.

\bibitem{Banks:1996my}
T.~Banks and N.~Seiberg, {\it Strings from matrices},  {\em Nucl. Phys.} {\bf
  B497} (1997) 41--55, [\href{http://xxx.lanl.gov/abs/hep-th/9702187}{{\tt
  hep-th/9702187}}].

\bibitem{Dijkgraaf:1997vv}
R.~Dijkgraaf, E.~P. Verlinde, and H.~L. Verlinde, {\it Matrix string theory},
  {\em Nucl. Phys.} {\bf B500} (1997) 43--61,
  [\href{http://xxx.lanl.gov/abs/hep-th/9703030}{{\tt hep-th/9703030}}].

\bibitem{Fischler:1997kp}
W.~Fischler, E.~Halyo, A.~Rajaraman, and L.~Susskind, {\it The incredible
  shrinking torus},  {\em Nucl. Phys.} {\bf B501} (1997) 409--426,
  [\href{http://xxx.lanl.gov/abs/hep-th/9703102}{{\tt hep-th/9703102}}].

\bibitem{Govindarajan:1997iw}
S.~Govindarajan, {\it Heterotic {M}(atrix) theory at generic points in {N}arain
  moduli space},  {\em Nucl. Phys.} {\bf B507} (1997) 589--608,
  [\href{http://xxx.lanl.gov/abs/hep-th/9707164}{{\tt hep-th/9707164}}].

\bibitem{Cachazo:2000ey}
F.~A. Cachazo and C.~Vafa, {\it Type {I}' and real algebraic geometry},
  \href{http://xxx.lanl.gov/abs/hep-th/0001029}{{\tt hep-th/0001029}}.

\bibitem{Ganor:1996zk}
O.~J. Ganor, S.~Ramgoolam, and W.~Taylor, {\it Branes, fluxes and duality in
  {M}(atrix)-theory},  {\em Nucl. Phys.} {\bf B492} (1997) 191--204,
  [\href{http://xxx.lanl.gov/abs/hep-th/9611202}{{\tt hep-th/9611202}}].

\bibitem{future}
 O. Aharony, Z. Komargodski and A. Patir, work in progress.

\bibitem{Hanany:2000fq}
A.~Hanany and B.~Kol, {\it On orientifolds, discrete torsion, branes and {M}
  theory},  {\em JHEP} {\bf 06} (2000) 013,
  [\href{http://xxx.lanl.gov/abs/hep-th/0003025}{{\tt hep-th/0003025}}].

\bibitem{Ho:1997yk}
P.~M. Ho, Y.~Y. Wu, and Y.~S. Wu, {\it Towards a noncommutative geometric
  approach to matrix compactification},  {\em Phys. Rev.} {\bf D58} (1998)
  026006, [\href{http://xxx.lanl.gov/abs/hep-th/9712201}{{\tt
  hep-th/9712201}}].

\end{thebibliography}\endgroup
\end{document}